\documentclass[sigconf]{aamas}

\usepackage{balance} 

\setcopyright{ifaamas}
\acmConference[AAMAS '26]{Proc.\@ of the 25th International Conference
on Autonomous Agents and Multiagent Systems (AAMAS 2026)}{May 25 -- 29, 2026}
{Paphos, Cyprus}{C.~Amato, L.~Dennis, V.~Mascardi, J.~Thangarajah (eds.)}
\copyrightyear{2026}
\acmYear{2026}
\acmDOI{}
\acmPrice{}
\acmISBN{}

\acmSubmissionID{240}

\title[AAMAS-2026 Formatting Instructions]{Agents of Diffusion: Enhancing Diffusion Language Models with Multi-Agent Reinforcement Learning for Structured Data Generation (Extended Version)}

\author{Aja Khanal}
\affiliation{
  \institution{University of Western Ontario}
  \city{London}
  \country{Canada}}
\email{akhanal3@uwo.ca}

\author{Kaushik T. Ranade}
\affiliation{
  \institution{University of Western Ontario}
  \city{London}
  \country{Canada}}
\email{kranade@uwo.ca}

\author{Rishabh Agrawal}
\affiliation{
  \institution{University of Western Ontario}
  \city{London}
  \country{Canada}}
\email{ragrawa9@uwo.ca}

\author{Kalyan S. Basu}
\affiliation{
  \institution{ICASSSD}
  \city{Cambridge}
  \country{Canada}}
\email{ks.basu@gmail.com}

\author{Apurva Narayan}
\affiliation{
  \institution{University of Western Ontario}
  \city{London}
  \country{Canada}}
\email{apurva.narayan@uwo.ca}

\begin{abstract}
Generating high-quality structured data such as JSON records, remains a fundamental challenge for large language models (LLMs), particularly when semantic richness must coexist with strict schema adherence. While autoregressive LLMs offer strong structural consistency, they often struggle with semantic variation and output diversity. In contrast, diffusion language models (DLMs) introduce powerful mechanisms for semantic richness and bidirectional decoding, yet lack the inductive biases needed for reliable structure preservation. We present \textbf{Agents of Diffusion (AoD)}, a novel framework that unifies the generative flexibility of DLMs with the reasoning capabilities of autoregressive models through language-mediated reinforcement learning. AoD frames structured text generation as a multi-agent alignment process, where a prompt optimization agent collaborates with a judge agent to iteratively guide a DLM using natural language feedback. This approach enables controllable, schema-consistent generation without modifying model parameters or relying on handcrafted constraints. AoD advances the state of controllable generation by demonstrating that diffusion models, when supervised by cooperative agents, can achieve both high semantic novelty and structural fidelity. Across multiple structured data benchmarks, AoD consistently outperforms diffusion and autoregressive baselines, establishing a new path forward for structure-aware, diversity-enhanced text synthesis. Code: https://github.com/Idsl-group/AgentsOfDiffusion.
\end{abstract}


\newcommand{\BibTeX}{\rm B\kern-.05em{\sc i\kern-.025em b}\kern-.08em\TeX}

\usepackage{algorithm}
\usepackage{algpseudocode}
\usepackage{tikz}
\usetikzlibrary{arrows.meta, positioning, shapes.geometric, calc}
\usepackage{xcolor}
\usepackage{amsthm}
\usepackage{multirow}
\usepackage[most]{tcolorbox} 
\usepackage{listings}        
\usepackage{xcolor}          
\usepackage{newfloat}
\tcbuselibrary{listings,skins,breakable} 

\theoremstyle{plain}
\newtheorem{theorem}{Theorem}

\newtheorem{proposition}{Proposition}

\newtheorem{compactitem}{Compactitem}

\makeatletter
\renewcommand{\paragraph}{%
  \@startsection{paragraph}{4}{0pt}%
    {-1ex plus -.5ex minus -.2ex}
    {-0.8em}
    {\normalfont\normalsize\bfseries}%
}%
\makeatother

\begin{document}


\pagestyle{fancy}
\fancyhead{}


\maketitle 


\section{Introduction}
\textbf{Agents of Diffusion (AoD)} is a multi-agent reinforcement learning framework for controllable data generation that pairs the semantic richness of diffusion language models (DLMs) with the structural precision of autoregressive large language model (LLM) agents. AoD explores a unique idea: use natural language feedback to supervise a DLM without fine-tuning, handcrafted rules, or scalar reward shaping. Two LLM agents (a prompt optimizer and a judge) communicate through verbal feedback to iteratively refine prompts, steering the DLM toward schema-conformant, diverse outputs.

Autoregressive LLMs are widely used in synthetic data pipelines because their inductive biases favor structure and token order \cite{huang_2023_large, raffel_2019_exploring}, yet these same biases can constrain diversity and trigger repetition or hallucination \cite{ji_2022_survey, zhao_2023_a}. DLMs, in contrast, generate text by iteratively denoising sequences in a non-causal, bidirectional manner \cite{hoogeboom_2021_autoregressive, luo_2022_understanding}, which encourages broader semantic variation. However, they lack positional priors for format preservation, which makes them poorly suited for structure-sensitive tasks such as nested JSON synthesis \cite{lee_2025_a}. AoD is designed to combine these strengths while compensating for their weaknesses.

Recent advances in prompt tuning, reinforcement learning, and agent-based coordination have improved autoregressive controllability \cite{ouyang_2022_training, shinn_2023_reflexion, liu_2023_large}, but comparable methods remain largely unexplored for DLMs due to their recent emergence. AoD closes this gap by enabling agentic supervision of DLMs through verbal alignment alone. Our optimization loop is parameter-free: the frozen DLM (LLaDA-8B) \cite{nie_2025_large} never updates its weights, and the agents interact only through natural language, which supports interpretability and model-agnostic control. To realize this, we introduce a reinforcement learning algorithm that blends proximal policy optimization (PPO) and REINFORCE principles to optimize prompt updates using natural language feedback as a surrogate reward signal.

We evaluate AoD on four structured generation benchmarks that require semantic fluency and strict JSON schema adherence: \textbf{MultiWOZ}, \textbf{Super-NaturalInstructions}, \textbf{Self-Instruct}, and \textbf{TruthfulQA}. These datasets contain nested fields, varied schema formats, and diverse linguistic styles, creating a challenging testbed for structure-aware DLM control. Across this suite, AoD achieves the highest \textbf{Task Success Rate} $(0.79)$ and the lowest \textbf{Field Overlap} $(0.29)$, outperforming diffusion and autoregressive baselines while indicating valid, non-memorized outputs.

The architecture is effective, reproducible, and accessible. AoD supports local open-weight models such as LLaMA 3.1 8B, Qwen-3 8B, DeepSeek-R1 8B, and Gemma-2 9B, as well as proprietary API-based models including GPT-4.1, GPT-4.1 Mini, and GPT-4.1 Nano. This flexibility enables operation in GPU-constrained environments and high-performance cloud settings alike. All experiments use a mix of consumer-grade hardware and API endpoints, showing that AoD does not require specialized infrastructure to produce high-quality, controllable structured generation.

\noindent\textbf{Contributions.}
(1) We introduce \textbf{Agents of Diffusion}, the first multi-agent RL framework to guide DLMs using natural language.\\
(2) We propose an optimization loop where LLM agents iteratively refine prompts through verbal critique, achieving schema-aligned control without reward modeling.\\
(3) We demonstrate reproducible state-of-the-art results on JSON-based instruction synthesis across multiple structured datasets, establishing a foundation for controllable generation in symbolic, format-constrained domains.

\section{Background and Related Work}
\subsection{Structured Textual Synthetic Data}
Synthetic data is increasingly important in machine learning, particularly when real data is limited, sensitive, or costly~\cite{bauer_2024_comprehensive}. While LLMs have shown early success in freeform text generation~\cite{keskar_2019_ctrl, anabytavor_2019_not, ye_2022_zerogen}, generating high-quality \textit{structured} data such as tabular records or JSON outputs remains a major challenge~\cite{josifoski_2023_exploiting}. LLMs often produce outputs that are syntactically correct but hallucinate and repeat outputs when required to generate data under nested structures. ~\cite{veselovsky_2023_generating, yu_2023_large}. Prior solutions relied on pipelines involving validation modules or latent modeling~\cite{wu_2024_unigen, dekoninck_2023_controlled}, but these approaches are difficult to scale. Inspired by the success of diffusion models in vision~\cite{rombach_2022_highresolution, ramesh_2021_zeroshot}, recent work has explored their application to text~\cite{gong_2022_diffuseq, zhang_2023_mixedtype}, though structure control remains limited. Our work addresses this gap by using a diffusion language model to enhance the semantic diversity of synthetic data, while ensuring structural fidelity through continuous evaluation in a multi-agent reinforcement learning setup.

\subsection{Autoregressive Language Models}
Autoregressive language models generate text by predicting each token sequentially, a paradigm that supports strong contextual coherence and structural alignment~\cite{vaswani_2017_attention, brown_2020_language, touvron_2023_llama}. Their unidirectional decoding makes them effective for producing syntactically valid and schema-compliant outputs, such as JSON or tabular formats ~\cite{gemmateam_2024_gemma, jiang_2023_mistral}. However, their reliance on left-to-right generation often limits output diversity, reinforcing high-probability patterns and leading to generic or repetitive sequences~\cite{chowdhery_2022_palm}. While sampling strategies offer some relief, the inherent sequential bias of AR LMs constrains their generative flexibility ~\cite{deepseekai_2025_deepseekr1, openai_2023_gpt4}. This motivates our exploration of diffusion models, which enable bidirectional and more diverse generation, while retaining structure through multi-agent control.

\subsection{Diffusion Language Models}
Diffusion language models (DLMs) generate text through iterative denoising, enabling more flexible and semantically diverse generation than autoregressive approaches. Diffusion-LM~\cite{li_2022_diffusionlm} introduced a continuous latent framework for controllable text synthesis, while LLaDA~\cite{nie_2025_large} extended masked-sequence diffusion to match or outperform AR baselines on reasoning and language tasks. DiffLM~\cite{zhou_2024_difflm} applied discrete diffusion to structured tabular data. While these models show promise, they face key limitations: structure preservation remains brittle in constrained formats like JSON, the effects of prompt tuning on DLM behavior are poorly understood, and their role in multi-agent coordination is largely unexplored. We address these gaps by embedding a frozen DLM in a multi-agent reinforcement learning loop with autoregressive LLM agents, using natural language feedback and prompt-space optimization to enable schema-consistent, semantically aligned generation.

\subsection{Prompt Tuning and Optimization}
Prompt optimization is widely used to adapt LLMs without fine-tuning, with methods ranging from hand-crafted reasoning strategies like Chain-of-Thought~\cite{wei_2022_chain} to automated approaches such as Promptbreeder~\cite{fernando_2023_promptbreeder} and EvoPrompt~\cite{guo_2023_connecting}. There equally exists self-rubric strategies for enhancing prompting as seen in CodecLM ~\cite{wang_2024_codeclm}. While effective, these methods treat prompt design as a static search problem, optimizing prompts offline without considering real-time feedback or generation dynamics ~\cite{Beurer-Kellner_Fischer_Vechev_2024}. This limits their adaptability, especially in tasks requiring structural precision or iterative refinement ~\cite{Dong_Ruan_Cai_Lai_Xu_Zhao_Chen_2025}. Our work reframes prompt optimization as a reinforcement learning problem, where a prompt agent learns to refine instructions based on natural language feedback. This approach is uniquely compatible with diffusion language models, whose iterative decoding allows prompts to guide generation over multiple denoising steps, enabling responsive and structure-aware control.

\subsection{Multi-Agent Language-Based Coordination}
Multi-agent systems (MAS) using LLMs have demonstrated strong performance on complex tasks by distributing reasoning across specialized, role-defined agents that communicate via natural language~\cite{guo_2024_large, tran_2025_multiagent}. Decentralized protocols such as critique, collaboration, or voting, enable robust, diverse solutions in domains like code synthesis and scientific discovery~\cite{wu_2023_autogen, du_2023_improving}. However, language-based coordination suffers from interaction drift, inconsistent agent behavior, and weak memory retention in multi-turn exchanges~\cite{cao_2024_survey, pan_2024_agentcoord}. We address these challenges by anchoring agent interactions around a diffusion language model, whose iterative and bidirectional decoding creates a stable shared reference point. This structure reduces drift, enforces semantic consistency, and grounds prompt refinement and evaluation across dialogue turns, improving coordination without requiring external memory modules or supervision.

\section{Methodology}
\begin{figure*}[t]
    \centering
    \includegraphics[width=\textwidth, trim=10 50 20 40, clip]{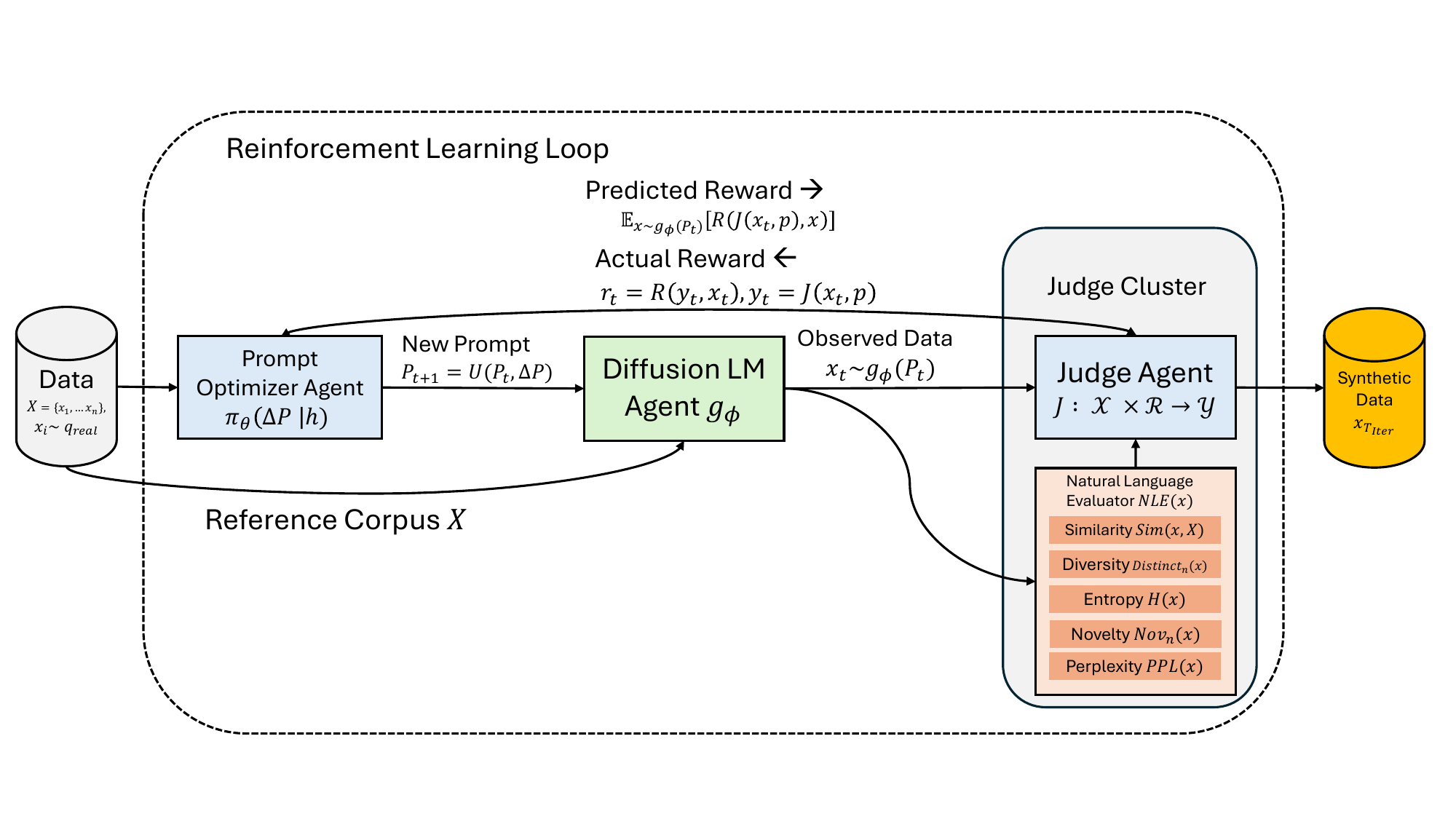}
    \caption{Agents of Diffusion: Overview of the multi-agent training framework.}
    \label{fig:aod}
\end{figure*}

\subsection{Preliminaries}
Let $\mathcal{X}$ be the space of structured text strings (JSON objects), and let $q_{\mathrm{real}}$ denote an unknown probability distribution over $\mathcal{X}$. We observe $n$ independent and identically distributed (i.i.d.) samples $X=\{x_1,\ldots,x_n\}$ with $x_i\sim q_{\mathrm{real}}$. A schema $S$ defines a valid subset $\mathcal{L}(S)\subseteq\mathcal{X}$, and a binary validator $V_S:\mathcal{X}\to\{0,1\}$ returns $V_S(x)=1$ if $x\in\mathcal{L}(S)$ and $0$ otherwise. Next, let $\mathcal{P}$ be the prompt space and $\mathcal{P}_\Delta$ the set of allowable prompt edits. A diffusion language model (DLM) with parameters $\phi$ defines reverse-time Markov kernels $g_{\phi,\tau}(z_{\tau-1}\mid z_\tau,P)$ for $\tau=1,\ldots,T$ and a noise prior $\nu_T$ on a latent token space $\mathcal{Z}$ (identified with $\mathcal{X}$ after decoding). To sample $x\sim g_\phi(P)$, one draws $z_T\sim\nu_T$, then iteratively denoises via $z_{\tau-1}\sim g_{\phi,\tau}(z_{\tau-1}\mid z_\tau,P)$ until $x=z_0$. The forward corruption process, defined by kernels $k_\tau(z_\tau\mid z_{\tau-1})$, is standard and omitted here.

The framework includes four interacting components $\mathcal{A}=\{\text{DLM},J,R,\pi\}$. The DLM $g_\phi$ produces text samples. The judge $J:\mathcal{X}\times\mathcal{R}\to\mathcal{Y}$ maps a candidate $x$ and rubric $\rho\in\mathcal{R}$ to feedback $y=J(x,\rho)\in\mathcal{Y}$. The scorer $R:\mathcal{Y}\times\mathcal{X}\to\mathbb{R}\times\mathbb{R}^k$ returns a scalar reward $r=R(y,x)$ and a subreward vector $\mathbf{s}=R_{\mathrm{vec}}(y,x)=(s_1,\ldots,s_k)$, where each $s_i$ captures a measurable quality such as semantic similarity, diversity, novelty, perplexity, or entropy.

The prompt optimizer is a stochastic policy $\pi_\theta(\Delta P\mid h)$ parameterized by $\theta\in\Theta\subset\mathbb{R}^d$, where $\Theta$ is the set of admissible parameter vectors (e.g., neural-network weights). Given a history $h\in\mathcal{H}$ summarizing previous prompts, samples, feedback, and scores, $\pi_\theta$ outputs a distribution over edits $\Delta P\in\mathcal{P}_\Delta$. The edit operator $U:\mathcal{P}\times\mathcal{P}_\Delta\to\mathcal{P}$ then deterministically updates prompts as $P^{+}=U(P,\Delta P)$. Outer optimization iterations are indexed by $t=1,\ldots,T_{\mathrm{iter}}$, while diffusion steps are indexed by $\tau=1,\ldots,T$.

Finally, for any random variable $Z$, $\mathbb{E}[Z]$ denotes expectation with respect to the randomness from $q_{\mathrm{real}}$, the DLM sampling process $g_\phi$, and the policy $\pi_\theta$, unless specified otherwise.

\subsection{Problem Formulation}
As shown in Figure ~\ref{fig:aod}, AoD formalizes the synthesis of textual JSON data as a controllable structured generation problem within a multi-agent reinforcement learning framework. The objective is to integrate the semantic diversity of diffusion language models with the structural precision of autoregressive agents. In this setup, a diffusion model proposes diverse JSON candidates, while two LLM agents iteratively refine the conditioning prompts through natural-language feedback to ensure schema conformity and semantic consistency. This cooperative feedback loop enables controllable generation of structured text without modifying model parameters.

At each outer iteration $t \in \{1,\ldots,T_{iter}\}$, the system maintains a prompt $P_t \in \mathcal{P}$ and a summary history $h_t \in \mathcal{H}$. 
The DLM $g_\phi$ generates a structured candidate $x_t \sim g_\phi(P_t)$. 
The judge $J$ evaluates $x_t$ under rubric $\rho \in \mathcal{R}$, producing feedback $y_t = J(x_t, \rho)$, and the scorer $R$ converts this into numerical signals $r_t = R(y_t, x_t)$ and $\mathbf{s}_t = R_{vec}(y_t, x_t)$. 
The prompt optimizer, parameterized by $\theta$, samples an edit $\Delta P_t \sim \pi_\theta(\Delta P \mid h_t)$ and updates the prompt via

\begin{equation}
    P_{t+1} = U(P_t, \Delta P_t).
\end{equation}

This defines an episodic Markov decision process (MDP) with state $z_t = (P_t, h_t)$, action $a_t = \Delta P_t$, transition kernel induced by $(g_\phi, J, R, U)$, and reward $r_t$. 
The policy parameters $\theta$ are optimized to maximize the expected discounted return:

\begin{equation}
    \max_{\theta}\;
    \mathbb{E}\!\left[\sum_{t=1}^{T_{iter}}\gamma^{t-1}r_t\right]
\end{equation}

subject to the resource constraints
$
Tokens \le B_{tok},\;
Calls \le B_{calls},\;
T_{iter} \le B_{iter},\;
\gamma \in [0,1).
$

At convergence ($T_{iter}$), performance is evaluated using terminal objectives that capture structural validity, semantic relevance, and diversity:
\[
\max_{\theta}\;
\alpha\,\mathbb{E}[V_S(x_{T_{iter}})]
+\beta\,\mathbb{E}[Sim(x_{T_{iter}},X)]
+\delta\,\mathbb{E}[Distinct_n(x_{T_{iter}})],
\]
or equivalently in constrained form,

\begin{equation}
    \max_{\theta}\; \mathbb{E}[Distinct_n(x_{T_{iter}})]
\end{equation}

such that
$ 
\mathbb{E}[V_S(x_{T_{iter}})] \ge \tau_{valid},\;
\mathbb{E}[Sim(x_{T_{iter}},X)] \ge \tau_{sim}.
$

The diffusion parameters $\phi$ remain fixed; controllability arises solely through the autoregressive policy $\pi_\theta$ acting on prompts.

\subsection{DLM Integrated Multi-Agent RL}

\begin{algorithm}[t]
\caption{Multi-Agent Reinforcement Learning Loop in Agents of Diffusion (AoD)}
\label{alg:aod}
\begin{algorithmic}[1]
\State \textbf{Input:} schema $S$, rubric $\rho$, diffusion model $g_\phi$, policy $\pi_\theta$, edit operator $U$
\State \textbf{Initialize:} prompt $P_1 \in \mathcal{P}$, summary history $h_1 \in \mathcal{H}$
\For{$t = 1$ \textbf{to} $T_{iter}$}
    \State Sample structured candidate $x_t \sim g_\phi(P_t)$
    \State Judge provides feedback $y_t = J(x_t, \rho)$
    \State Compute reward $r_t = R(y_t, x_t)$ and subrewards $\mathbf{s}_t = R_{vec}(y_t, x_t)$
    \State Update history summary $h_t = f(P_t, x_t, y_t, \mathbf{s}_t)$
    \State Sample prompt edit $\Delta P_t \sim \pi_\theta(\Delta P \mid h_t)$
    \State Apply edit: $P_{t+1} = U(P_t, \Delta P_t)$
    \State Update policy parameters:
        $\theta \leftarrow \theta + \eta \, \widehat{\nabla}_\theta \, \mathbb{E}_{\pi_\theta}[r_t]$
\EndFor
\State \textbf{Return:} final prompt $P_{T_{iter}}$, final candidate $x_{T_{iter}}$
\end{algorithmic}
\end{algorithm}

\paragraph{Prompt Optimization Agent.}
The prompt optimization agent governs controllability in text-based structured JSON synthesis by steering the diffusion language model toward schema-valid, semantically coherent generations. It is instantiated as an autoregressive large language model because prompt edits are sequential and token-dependent, making autoregressive architectures ideal for learning discrete edit trajectories in text space. The agent defines a stochastic policy $\pi_\theta(\Delta P \mid h)$ parameterized by $\theta$, generating contextually guided prompt updates that modulate the conditional output distribution of $g_\phi$. Its objective is to maximize the expected reward provided by the judge–scorer pair $(J,R)$:
\begin{equation}
\label{eq:aod_policy_objective}
\pi_\theta^* = \arg\max_{\pi_\theta}\, \mathbb{E}_{x \sim g_\phi(P)}[\,R(J(x,\rho),x)\,],
\end{equation}
where $\rho$ is a task-specific rubric capturing schema and semantic fidelity. Since $g_\phi$ is non-differentiable, $\pi_\theta$ serves as a surrogate functional optimizer that approximates the gradient of $\mathcal{T}(P)=\mathbb{E}_{x\sim g_\phi(P)}[R(J(x,\rho),x)]$ through discrete, language-conditioned updates rather than backpropagation.

\begin{theorem}
\label{thm:aod_convergence}
Let $\mathcal{T}(P)=\mathbb{E}_{x\sim g_\phi(P)}[R(J(x,\rho),x)]$ as in (\ref{eq:aod_policy_objective}). 
If $\mathcal{T}(P)$ is locally Lipschitz and prompt edits $\Delta P_t$ sampled from $\pi_\theta(\Delta P \mid h_t)$ are bounded, then the iterative update 
$P_{t+1}=U(P_t,\Delta P_t)$ constitutes a contraction mapping in expectation for sufficiently small step size. 
Thus, the sequence $\{P_t\}$ converges to a fixed point $P^*$ satisfying 
\[
\mathcal{T}(P^*) = \max_P \mathcal{T}(P),
\]
ensuring stable convergence toward reward-aligned, schema-consistent prompts. 
\end{theorem}

This agent operationalizes discrete autoregressive reasoning as a control layer over diffusion dynamics, transforming natural-language feedback into token-level schema alignment steps. For instance, under a booking schema $S$ with fields \texttt{departure\_city}, \texttt{arrival\_city}, and \texttt{date}, an initial prompt such as “Generate travel details” may elicit feedback like “Use JSON format and include all fields.” Through iterative updates $\Delta P_t$, the agent refines this into \texttt{"Generate a JSON object with fields \{origin, destination, date\} using YYYY-MM-DD format."} Each refinement incrementally aligns the DLM’s conditional distribution $p_\phi(x \mid P_t)$ with the valid schema subset $\mathcal{L}(S)$ while preserving semantic diversity. Theoretically, this mechanism bridges discrete symbolic reasoning and stochastic generation, enabling reinforcement-driven adaptation in structured text synthesis.

\paragraph{Diffusion Language Model Agent.}
The diffusion language model (DLM) $g_\phi(z_{\tau-1} \mid z_\tau, P)$ serves as the generative backbone for synthesizing text-based structured JSON data. It defines a reverse-time Markov process that reconstructs text tokens from gradually denoised latent representations, yielding the conditional distribution $p_\phi(x \mid P)$. Unlike autoregressive models that factorize $p(x \mid P)$ sequentially as $\prod_i p(x_i \mid x_{<i}, P)$, the DLM estimates $p_\phi(x \mid P)$ implicitly through iterative denoising. This non-causal, bidirectional formulation allows each denoising step to condition on global context rather than local token dependencies, resulting in broader coverage of valid schema-conformant configurations.

\begin{proposition}
Let $p_\phi(x \mid P)$ and $p_{AR}(x \mid P)$ denote diffusion and autoregressive conditional distributions trained on the same structured dataset with schema $S$. If both minimize divergence from the real data distribution $q_{\mathrm{real}}(x)$ under bounded reconstruction error and finite diffusion horizon $T$, then
\[
\mathrm{KL}(q_{\mathrm{real}}(x)\,\|\,p_\phi(x \mid P)) \le \mathrm{KL}(q_{\mathrm{real}}(x)\,\|\,p_{AR}(x \mid P)),
\]
indicating that diffusion better approximates the real data manifold and captures a wider set of semantically valid configurations. 
\end{proposition}

Theoretically, this property positions the DLM as the diversity-preserving agent within AoD. It expands the support of $p_\phi(x \mid P)$ across multiple valid schema realizations, enabling the synthesis of varied yet coherent JSON structures. The autoregressive agents, in contrast, provide the constraint mechanism that ensures syntactic and semantic adherence to $\mathcal{L}(S)$. Together, they form a complementary system: diffusion drives diversity, while autoregression enforces structure. This achieves controlled, schema-aligned generation of structured text data.

\paragraph{Judge Agent.}
The judge cluster combines the LLM-based judge $J$ and the Natural Language Evaluator (NLE), forming the evaluation subsystem responsible for interpreting and supervising outputs from the diffusion language model $g_\phi$. The NLE receives each generated JSON sample $x \in \mathcal{X}$ and computes five quantitative metrics defined in the preliminaries: semantic similarity $\mathrm{Sim}(x,X)$, diversity $\mathrm{Distinct}\text{-}n(x)$, entropy $H(x)$, novelty $\mathrm{Nov}\text{-}n(x)$, and perplexity $\mathrm{PPL}(x)$. These metrics jointly describe how well $x$ aligns with the reference dataset $X$, how varied and fluent it is, and whether it generalizes beyond seen examples. The NLE then converts these numeric values into structured natural language statements, which summarize deviations and attributions in an interpretable form. The LLM judge $J$, instantiated as an autoregressive model, consumes this structured textual feedback together with the rubric $\rho$ and evaluates each sample through a fixed set of rubric-aligned yes/no questions (e.g., ``Is the JSON structurally complete?'', ``Are all required fields present?'', ``Is the text semantically faithful?''). Based on both the quantitative assessments and its own contextual reasoning, $J$ generates the final critique $y = J(x,\rho)$, which the scorer translates into a scalar reward $r = R(y,x)$ and subreward vector $\mathbf{s} = R_{\text{vec}}(y,x)$. For example, the cluster might produce: “\textit{The JSON is fluent but missing \texttt{arrival\_city}; the date format should be YYYY-MM-DD.}” This feedback then guides the prompt optimization policy $\pi_\theta(\Delta P \mid h)$, completing the reinforcement loop.  

 Correspondingly, the NLE isolates measurable properties of the generated text, providing low-variance, disentangled feedback signals that prevent noisy gradients and improve the reliability of downstream optimization. The LLM judge $J$ transforms these discrete measurements into a smooth, natural-language surrogate of the underlying reward landscape, enabling gradient-free optimization while maintaining semantic transparency. The autoregressive formulation of $J$ is particularly important for structured JSON synthesis, as evaluating conformance, field ordering, and key dependencies requires sequential reasoning over tokens. By processing feedback in a left-to-right manner, $J$ preserves causal consistency in its critiques and ensures alignment with how $\pi_\theta$ performs token-level edits to $P$. Thus, autoregressive reasoning enforces structured coherence while maintaining semantic flexibility, allowing the system to generalize across diverse schemas without manual rule engineering.  

Theoretically, the cluster defines an expected feedback operator 

\begin{equation}
    \mathcal{T}(P)=\mathbb{E}_{x\sim g_\phi(P)}[R(J(x,\rho),x)]
\end{equation}

that stabilizes learning in the non-differentiable environment $(g_\phi,J,R)$. The NLE grounds $\mathcal{T}(P)$ in verifiable quantitative signals, while the autoregressive judge smooths discontinuities by mapping discrete validation outcomes to continuous linguistic explanations. This synergy reduces variance in the policy-gradient estimate, improves credit assignment, and enforces reward monotonicity with respect to schema-conformant and semantically faithful outputs. Together, they form a theoretically consistent bridge between numeric supervision and symbolic prompt control, balancing the structured precision of autoregressive models with the generative diversity of diffusion-based synthesis.

\begin{proposition}
Let $\mathcal{T}(P)=\mathbb{E}_{x\sim g_\phi(P)}[R(J(x,\rho),x)]$. Suppose each component in $R_{\text{vec}}(y,x)$ is bounded, the mapping from the NLE’s metric vector to textual feedback $y=J(x,\rho)$ is Lipschitz with constant $L_J$ with respect to $(\mathrm{Sim}(x,X),\mathrm{Distinct}\text{-}n(x),H(x),\mathrm{Nov}\text{-}n(x),\mathrm{PPL}(x))$, and $R$ is monotone in these components. Then $\mathcal{T}(P)$ is locally Lipschitz in $P$ and preserves ordering with respect to semantic fidelity: if $P_1,P_2$ induce samples such that $\mathbb{E}[\mathrm{Sim}(x,X)\mid P_1]>\mathbb{E}[\mathrm{Sim}(x,X)\mid P_2]$, then $\mathcal{T}(P_1)>\mathcal{T}(P_2)$. Consequently, policy updates driven by $\mathcal{T}(P)$ are stable in expectation and prioritize semantically faithful prompts.
\end{proposition}

\section{Experiments and Results}
\subsection{Experimental Setup}

\paragraph{Hardware. }
To demonstrate the accessibility and reproducibility of AoD, all experiments were run on a consumer-grade workstation with an AMD Ryzen 9 7900X (12-core, 24-thread, 4.7 GHz base), 32 GB DDR5 RAM, and an NVIDIA RTX 4080 SUPER GPU (16 GB VRAM). This setup reflects hardware that is widely available to individual researchers and developers, ensuring that AoD does not rely on specialized infrastructure or large-scale compute clusters.

\paragraph{Models. }
Our experiments use both open-source and API-based language models to highlight AoD's flexibility and hardware independence. Eight autoregressive models were used for the prompt optimizer and LLM judge roles: \textbf{LLaMA-3.1 8B} (32 layers, 40 heads, 4-bit quantization; temperature 0.7 for prompting, 0.2 for judgment), \textbf{Qwen-3 8B} (multilingual, LoRA-enabled, nucleus sampling with $p=0.9$), \textbf{DeepSeek-R1 8B} (NTK-aware, top-$k$ sampling with $k=40$), \textbf{Gemma-2 9B} (beam search width 3), \textbf{Mistral 7B} (grouped-query attention, 8-bit decoding), and three API-based models: \textbf{GPT-4.1 Nano}, \textbf{Mini}, and \textbf{GPT-4.1}, which support lightweight to high-fidelity generation. All experiments used the same autoregressive model for both the prompt optimizer and LLM judge for consistency and computational simplicity. The generator agent, by contrast, was exclusively \textbf{LLaDA 8B}, a discrete diffusion language model with 32 layers, sinusoidal embeddings, 1024-token input, and $T=12$ denoising steps. LLaDA operates in FP16 mode and disables classifier-free guidance to preserve sampling diversity. Across both local and API models, AoD demonstrates consistent, high-quality performance without reliance on specialized hardware.

\paragraph{Datasets. }
We evaluate our framework to synthesize JSON data on four publicly available datasets chosen for their diversity in structure, semantics, and generation objectives. \textbf{MultiWOZ}~\cite{budzianowski_2018_multiwoz} is a multi-domain dialogue dataset with rich slot-filling annotations, useful for testing structure preservation and schema alignment. \textbf{Super-NaturalInstructions}~\cite{wang_2022_supernaturalinstructions} contains diverse instruction-response pairs across hundreds of tasks, enabling generalization over prompt types. \textbf{TruthfulQA}~\cite{lin_2021_truthfulqa} provides factuality-challenging questions, useful for evaluating hallucination and semantic precision. \textbf{Self-Instruct}~\cite{wang_2022_selfinstruct} consists of instruction-based examples distilled from LLMs, supporting experiments on prompt-response synthesis in alignment-critical tasks. Each dataset is randomly subsampled prior to training and evaluation to reduce computational overhead and mitigate memorization. By exposing the model to only a small, randomly selected portion of the full dataset, we minimize the risk of copying specific examples during generation and ensure that performance reflects generalization to unseen instructions.

\paragraph{Baselines. }
We compare AoD against six complementary baselines that represent the main paradigms of structured text generation. 
\textbf{Diffusion-LM}~\cite{li_2022_diffusionlm} and \textbf{DiffLM}~\cite{zhou_2024_difflm} are diffusion-based models emphasizing semantic diversity and multimodal coverage, serving as diversity-oriented references. 
\textbf{CodecLM}~\cite{wang_2024_codeclm}, \textbf{PromptBreeder}~\cite{fernando_2023_promptbreeder}, and \textbf{EvoPrompt}~\cite{guo_2023_connecting} are autoregressive prompt-optimization methods that provide strong control-oriented baselines, reflecting existing strategies for structured prompting and language-driven refinement. 
Finally, \textbf{UniGen}~\cite{wu_2024_unigen} embodies validation-based synthesis, enforcing symbolic and schema-level constraints at generation time. 
Together, these six baselines span the design space of \textit{structure control} (AR-based), \textit{diversity} (diffusion-based), and \textit{constraint enforcement} (validation-based), providing a comprehensive comparison framework for AoD’s contribution: unifying all three through a multi-agent reinforcement-learning mechanism that produces schema-conformant yet semantically diverse structured JSON data.

\paragraph{Evaluation Metrics.}
Our evaluation framework separates metrics used for agent feedback from those used for independent verification. During training, the judge agent leverages five grounded metrics: perplexity $\mathrm{PPL}(x)$, semantic similarity $\mathrm{Sim}(x,X)$, diversity $\mathrm{Distinct}\text{-}n(x)$, token entropy $H(x)$, and novelty $\mathrm{Nov}\text{-}n(x)$, to generate interpretable natural language feedback for the prompt optimizer. These metrics provide structural and semantic supervision without exposing scalar reward values directly, thereby avoiding reinforcement bias or overfitting to numeric targets. However, as a measure for independent evaluation, we report standard text-based quality metrics including BLEU, ROUGE, and METEOR to quantify syntactic and lexical correspondence between generated and reference samples. This ensures that our reported performance reflects true generation quality rather than reinforcement feedback bias. Furthermore, to assess the downstream finetunability and functional reliability of the synthesized data, we also compute the \textit{Task Success Rate} (TSR), which measures the proportion of valid, semantically consistent, and diverse generations meeting all constraints.

\paragraph{Memorization and Collusion Verification.}
The five reward metrics used during training (perplexity $\mathrm{PPL}(x)$, semantic similarity $\mathrm{Sim}(x,X)$, diversity $\mathrm{Distinct}\text{-}n(x)$, token entropy $H(x)$, and novelty $\mathrm{Nov}\text{-}n(x)$) jointly regulate memorization and collusion within the multi-agent reinforcement learning loop. Each metric enforces a distinct behavioral constraint: $\mathrm{PPL}(x)$ ensures linguistic coherence and penalizes degenerate text, $\mathrm{Sim}(x,X)$ promotes semantic alignment, $\mathrm{Distinct}\text{-}n(x)$ and $H(x)$ encourage lexical variability, and $\mathrm{Nov}\text{-}n(x)$ penalizes verbatim reuse of the reference dataset $X$. The judge uses these quantitative signals to answer a fixed set of rubric-based yes/no questions, ensuring that reinforcement is grounded in objective structure and meaning rather than hidden coordination between agents. Because the prompt optimizer $\pi_\theta$ never directly observes scalar rewards but instead receives natural language feedback derived from them, it cannot exploit the reward function through collusion or memorization. Numeric trends across iterations provide diagnostic signals, such as a simultaneous increase in $\mathrm{Sim}(x,X)$ and decrease in $\mathrm{Nov}\text{-}n(x)$, which explicitly reveal potential leakage or overfitting.

To further ensure independence and verify that generated data remains distinct from the training corpus, we introduce the \textit{Field Overlap} metric as a post-hoc measure. Field Overlap computes the proportion of key–value pairs or fields in generated JSON samples that exactly match those in the reference set. High overlap values indicate potential copying or memorization, while low overlap combined with low $\mathrm{PPL}(x)$ suggests faithful generalization with coherent generation. Unlike $\mathrm{Sim}(x,X)$, which captures semantic similarity, Field Overlap explicitly measures structural duplication, making it a direct test for memorization or cross-agent information leakage. Together, these signals enable both in-loop and independent verification of data novelty, ensuring that AoD produces diverse, semantically faithful, and unbiased synthetic JSON data.

\subsection{Discussion}

\begin{table*}[t]
\captionsetup{skip=2pt} 
\setlength{\tabcolsep}{4pt} 
\renewcommand{\arraystretch}{0.9} 
\setlength{\aboverulesep}{0pt}
\setlength{\belowrulesep}{0pt}
\setlength{\cmidrulesep}{0.2em}

\centering
\caption{Comparison of AoD with baselines, The first five metrics correspond to those used in training, while the last five serve as independent evaluation metrics. Higher is better for all metrics except Perplexity and Field Overlap. Values averaged across all datasets and Prompt Optimizer + Judge LLM pairs for AoD. Each experiment was repeated 15 times.}
\label{tab:main_results}

\resizebox{0.97\linewidth}{!}{
\begin{tabular}{@{}lccccc|ccccc@{}} 
\toprule
\textbf{Model} & \textbf{Similarity} & \textbf{Diversity} & \textbf{Novelty} & \textbf{Entropy} & \textbf{Perplexity} & \textbf{BLEU} & \textbf{ROUGE-L} & \textbf{METEOR} & \textbf{TSR} & \textbf{Field Overlap} \\
\midrule
\multicolumn{11}{c}{\small\textit{Static Autoregressive Baselines (single-pass prompting)}} \\
LLaMA-3.1 8B & 0.86 & 0.42 & 0.48 & 5.18 & 21.6 & 33.9 & 38.1 & 27.9 & 0.71 & 0.38 \\
Qwen-3 8B & 0.85 & 0.44 & 0.50 & 5.22 & 22.3 & 34.1 & 37.6 & 27.8 & 0.70 & 0.36 \\
DeepSeek-R1 8B & 0.87 & 0.41 & 0.47 & 5.14 & 20.9 & 35.2 & 38.8 & 28.0 & 0.73 & 0.35 \\
Gemma-2 9B & 0.84 & 0.43 & 0.49 & 5.19 & 22.0 & 34.5 & 37.8 & 28.1 & 0.72 & 0.37 \\
Mistral 7B & 0.83 & 0.44 & 0.48 & 5.25 & 23.2 & 33.7 & 37.2 & 27.5 & 0.69 & 0.39 \\
GPT-4.1 Nano & 0.84 & 0.46 & 0.55 & 5.16 & 21.9 & 30.5 & 36.0 & 26.2 & 0.66 & 0.38 \\
GPT-4.1 Mini & 0.84 & 0.47 & 0.56 & 5.12 & 21.5 & 30.9 & 36.5 & 26.4 & 0.67 & 0.37 \\
GPT-4.1 & 0.85 & 0.48 & 0.56 & 5.10 & 21.2 & 31.0 & 36.8 & 26.6 & 0.68 & 0.37 \\
\midrule
\multicolumn{11}{c}{\small\textit{Diffusion and Prompt-Based Baselines}} \\
Diffusion-LM~\cite{li_2022_diffusionlm} & 0.72 & 0.60 & 0.72 & 5.82 & 29.4 & 28.1 & 33.5 & 25.1 & 0.61 & 0.42 \\
DiffLM~\cite{zhou_2024_difflm} & 0.74 & 0.63 & 0.70 & 5.90 & 28.6 & 27.5 & 32.9 & 24.6 & 0.63 & 0.41 \\
UniGen~\cite{wu_2024_unigen} & 0.78 & 0.52 & 0.63 & 5.64 & 27.5 & 30.8 & 35.0 & 26.0 & 0.67 & 0.40 \\
PromptBreeder~\cite{fernando_2023_promptbreeder} & 0.80 & 0.51 & 0.59 & 5.51 & 25.7 & 31.2 & 36.7 & 26.5 & 0.68 & 0.38 \\
EvoPrompt~\cite{guo_2023_connecting} & 0.81 & 0.49 & 0.57 & 5.48 & 25.1 & 32.4 & 37.0 & 27.0 & 0.70 & 0.37 \\
CodecLM~\cite{wang_2024_codeclm} & 0.82 & 0.47 & 0.56 & 5.42 & 24.8 & 33.0 & 37.5 & 27.3 & 0.71 & 0.36 \\
LLaDA~\cite{nie_2025_large} & 0.79 & 0.69 & 0.81 & 6.03 & 27.0 & 29.5 & 34.2 & 25.8 & 0.69 & 0.35 \\
\midrule
\textbf{AoD (ours)} & \textbf{0.88} & \textbf{0.82} & \textbf{0.83} & \textbf{6.10} & \textbf{22.1} & \textbf{35.6} & \textbf{40.1} & \textbf{29.3} & \textbf{0.79} & \textbf{0.29} \\
\bottomrule
\end{tabular}
}
\end{table*}

Table~\ref{tab:main_results} highlights that AoD achieves a rare balance between structural precision and generative diversity, outperforming both diffusion-based and autoregressive systems. High Similarity (0.88) combined with strong Diversity (0.72) and Novelty (0.83) demonstrates that AoD generates data that remains semantically faithful while exploring new schema-consistent configurations. The Entropy score (6.10) indicates balanced lexical richness rather than repetitive phrasing, and the low Perplexity (22.1) confirms fluent and coherent language modeling. This pattern is not coincidental; it directly reflects the balance between exploration and regulation within AoD’s architecture. The diffusion generator introduces stochastic breadth, while the judge cluster applies linguistic and structural constraints that stabilize the output space. Reinforcement-guided optimization aligns these opposing forces, producing samples that are both creative and compositionally valid, even in datasets with complex nested structures.

Independent metrics reinforce this interpretation. AoD leads on BLEU, ROUGE-L, and METEOR, showing that its diversity does not compromise grammatical or semantic fidelity. The high Task Success Rate (0.79) indicates that generated records satisfy both content and structure requirements, while the lowest Field Overlap (0.29) confirms that AoD avoids memorization by generating distinct key-value combinations unseen in training. These numerical patterns arise from the multi-agent reward structure: the natural language evaluator (NLE) introduces interpretive continuity by converting discrete metric signals into graded linguistic feedback, while the autoregressive judge performs sequential validation across keys and fields. This layered supervision stabilizes reward propagation, making optimization smoother and preventing overfitting to numeric heuristics. The result is visible in the metrics—Similarity and Perplexity improve simultaneously, Diversity and Novelty rise without structural drift, and Entropy remains high yet coherent. AoD’s reinforcement signals therefore encode both form and meaning, producing generalization that persists across unseen distributions.

These dynamics also clarify why AoD achieves an uncommon combination of low Perplexity and high Entropy. Traditional diffusion systems increase diversity but often generate syntactically unstable text, while autoregressive systems enforce structure at the cost of variability. AoD bridges this divide by coupling diffusion-driven exploration with sequential constraint verification. The NLE’s linguistic grounding allows lexical and semantic expansion to occur in a controlled way, and the judge’s autoregressive reasoning enforces causal dependencies between fields. Together, they produce the observed equilibrium: high-entropy text that remains syntactically fluent and semantically consistent. This mechanism explains why AoD avoids the typical diffusion drift toward incoherence and the autoregressive bias toward repetition. The differences among baseline models further contextualize AoD’s superiority. Static autoregressive systems such as LLaMA and Qwen prioritize conditional likelihood maximization, maintaining high Similarity but collapsing on Diversity and Novelty due to deterministic decoding. Diffusion baselines like DiffLM or Diffusion-LM invert this pattern, producing diverse but structurally fragile data because their denoising trajectories lack schema-aware conditioning. Prompt-evolution frameworks like EvoPrompt and PromptBreeder improve variability through heuristic mutation but fail to sustain progress because they lack credit assignment across sequential edits. In contrast, AoD closes this optimization loop through dynamic prompt refinement driven by multi-dimensional feedback from the judge cluster. The prompt optimizer learns a policy that adjusts prompts not only based on reward magnitude but also on the linguistic context of errors, enabling continual improvement across iterations.

\begin{figure}[htbp]
    \includegraphics[width=\linewidth]{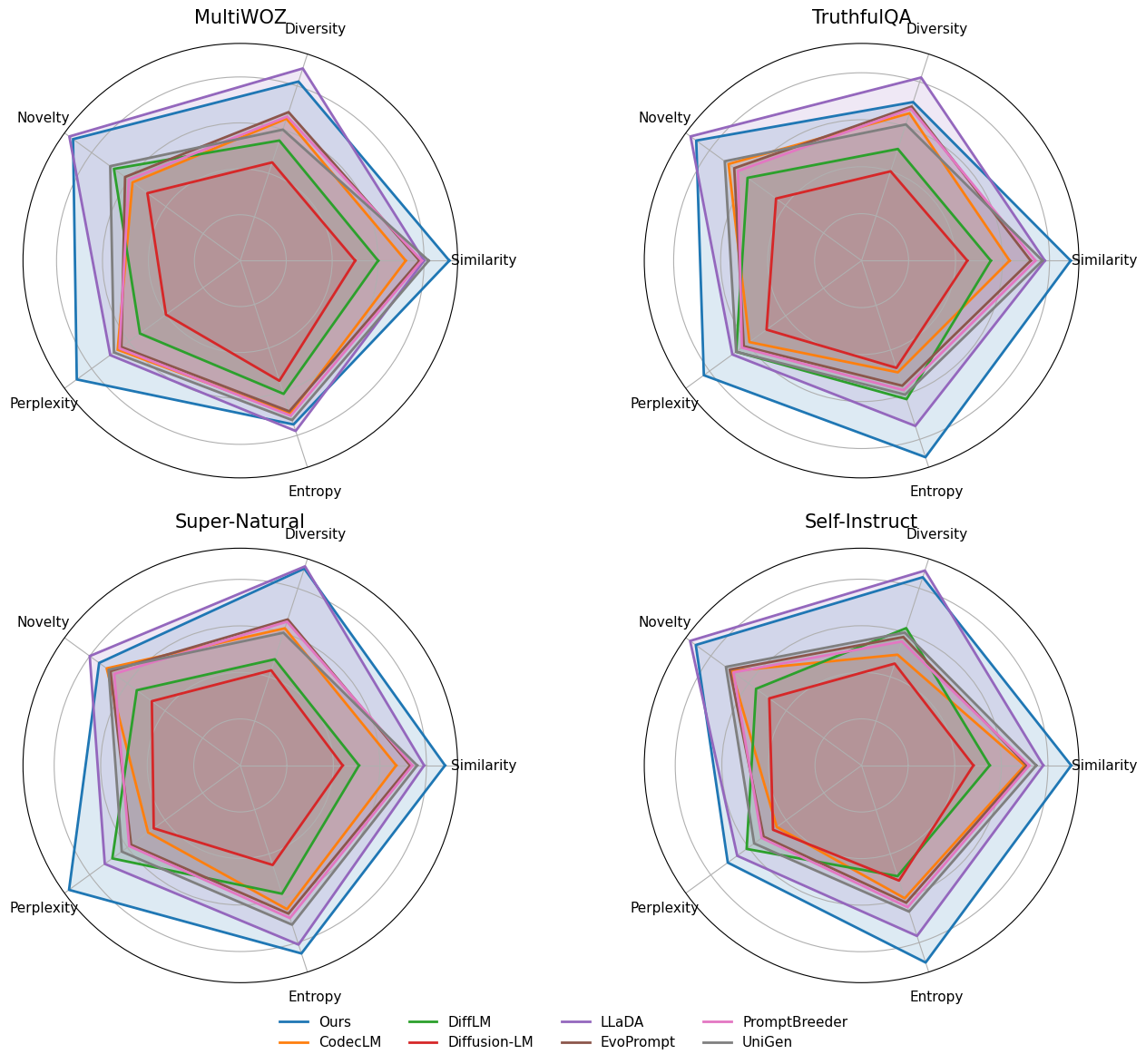}
    \caption{Comparison of normalized metrics across datasets.}
    \label{fig:comp}
\end{figure}

Figure~\ref{fig:comp} illustrates AoD’s consistency across datasets, where its polygons expand uniformly along all metric axes. In structurally complex domains like \textit{MultiWOZ} and \textit{Super-Natural}, AoD sustains high Similarity and low Perplexity while widening Diversity and Entropy, demonstrating its ability to preserve schema integrity while encouraging lexical variation. On reasoning-heavy datasets such as \textit{TruthfulQA} and \textit{Self-Instruct}, AoD balances Novelty and Similarity, showing that the judge cluster and diffusion generator collaboratively regulate generalization without overfitting. Competing methods either favor diversity at the expense of structure (Diffusion-LM, DiffLM) or maintain structure but exhibit reduced novelty (PromptBreeder, EvoPrompt). The near-regular shape of AoD’s region across all four datasets confirms stable performance and adaptability to different data modalities, reinforcing its role as a schema-faithful yet diverse structured data generator.

\subsection{Ablation Study}
\paragraph{Agentic and Reward Ablation.}

\begin{figure*}[t]
    \centering
    \includegraphics[width=\textwidth]{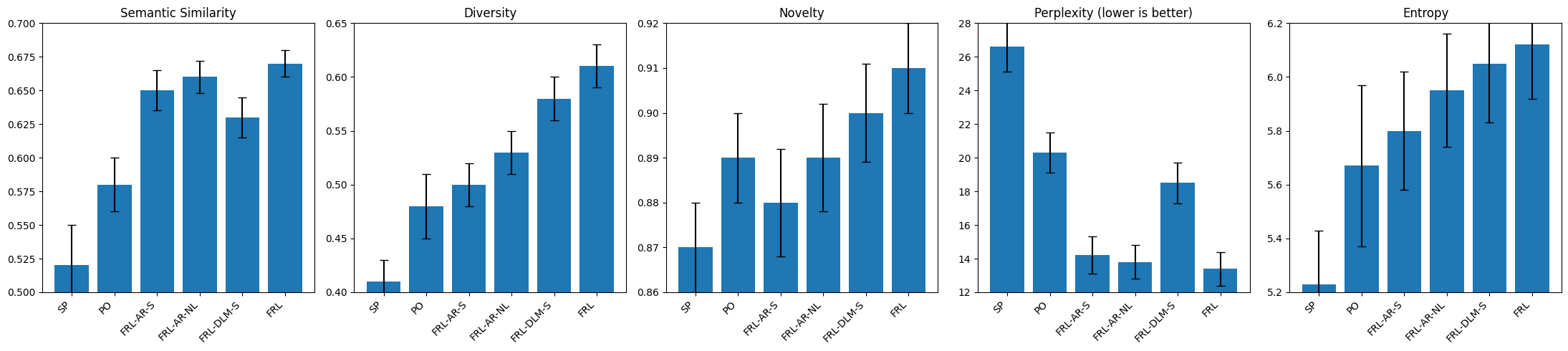}
    \caption{Ablation of agents, averaged across datasets and Optimizer–Judge pairs. SP = Static Prompt, PO = Prompt Optimizer, FRL–AR–S = Autoregressive generator with scalar rewards, FRL–AR–NL = Autoregressive generator with natural language feedback, FRL–DLM–S = Diffusion generator with scalar rewards, FRL = Diffusion generator with natural language feedback.}

    \label{fig:ablation}
\end{figure*}

Figure~\ref{fig:ablation} illustrates the impact of each agentic component and feedback mechanism within AoD. Introducing the autoregressive prompt optimizer (PO) leads to improvements across all metrics. This indicates that prompt refinement, even without reinforcement, enhances fluency and schema alignment by allowing structured edits over iterations. However, Diversity and Entropy remain constrained since optimization is deterministic and lacks stochastic exploration. Adding reinforcement learning with scalar rewards (\textit{FRL--AR--S}) further enhances Similarity and lowers Perplexity, as scalar feedback enables structured reward shaping for grammatical and syntactic correctness. Yet, gains in Diversity and Entropy are limited by the autoregressive generator’s decoding bias, which narrows the output space to high-likelihood continuations. Replacing scalar rewards with natural language (\textit{FRL--AR--NL}) yields smoother reward propagation and improved fluency, as the judge's interpretable feedback offers token-level supervision. Nevertheless, the sequential nature of the autoregressive generator continues to limit the attainable variety.

Transitioning to a diffusion generator while maintaining scalar rewards (\textit{FRL--DLM--S}) produces an increase in Diversity, Novelty, and Entropy. The diffusion-based generator $g_\phi$ enables bidirectional context propagation, allowing multiple semantic trajectories to emerge under the same prompt while maintaining structural coherence. However, in the absence of natural language feedback, Similarity improvements are modest and Perplexity variance rises, reflecting coarse reward alignment. The complete AoD configuration (\textit{FRL}) combines the strengths of both: diffusion-driven diversity with linguistically grounded feedback. The LLM Judge and NLE stabilize the learning dynamics, reducing Perplexity and reinforcing high Similarity and Entropy while maintaining elevated Novelty. 


\begin{table}[t]
\centering
\caption{Transferability for prompt optimizer (rows) and judge (columns). Reports Diversity (\textuparrow) and GPU runtime seconds (\textdownarrow) averaged for all datasets. LLaDA is fixed generator. L= LLaMA, Q = Qwen, D = DeepSeek, G = Gemma, M = Mistral.}
\label{tab:model_transfer_single}
\setlength{\tabcolsep}{3pt}
\begin{tabular}{lcccccccccc}
\toprule
& \multicolumn{2}{c}{L} & \multicolumn{2}{c}{Q} & \multicolumn{2}{c}{D} & \multicolumn{2}{c}{G} & \multicolumn{2}{c}{M} \\
\cmidrule(lr){2-3}\cmidrule(lr){4-5}\cmidrule(lr){6-7}\cmidrule(lr){8-9}\cmidrule(lr){10-11}
\textbf{Opt.} & Div. & Rt. & Div. & Rt. & Div. & Rt. & Div. & Rt. & Div. & Rt. \\
\midrule
L & 0.83 & 112 & 0.81 & 110 & 0.80 & 115 & 0.78 & 117 & 0.82 & 113 \\
Q & 0.82 & 118 & 0.85 & 119 & 0.84 & 116 & 0.79 & 118 & 0.80 & 121 \\
D & 0.80 & 115 & 0.83 & 117 & 0.86 & 114 & 0.82 & 118 & 0.81 & 120 \\
G & 0.79 & 120 & 0.81 & 121 & 0.82 & 122 & 0.84 & 118 & 0.83 & 119 \\
M & 0.82 & 111 & 0.83 & 113 & 0.81 & 114 & 0.79 & 115 & 0.84 & 112 \\
\bottomrule
\end{tabular}
\end{table}

\paragraph{Model Transferability.}

Table~\ref{tab:model_transfer_single} highlights AoD’s model-agnostic behavior across various combinations of autoregressive LLMs serving as the Prompt Optimizer and Judge agents, with LLaDA fixed as the generator. Across all pairings, Diversity scores remain consistently high (0.79–0.86), indicating that reinforcement-driven coordination generalizes regardless of the underlying model architecture. This demonstrates that AoD’s policy learning operates on the shared language space of feedback and prompts, rather than relying on any specific model’s internal representations. The prompt–feedback exchange mechanism $\pi_\theta(\Delta P \mid h)$ is thus invariant to the optimizer and judge configurations, enabling interchangeable agents without performance collapse. Runtime results further support AoD’s reproducibility on consumer-grade hardware. Average GPU runtimes per feedback–generation cycle range from 110–122 seconds, even for 8–9B parameter models, confirming that multi-agent rollouts remain tractable under mid-range configurations. This efficiency stems from the frozen generator $g_\phi$ and the lightweight communication loop between autoregressive agents, which limits backpropagation overhead. Together, these results establish that AoD can be replicated using open-weight or API-based LLMs while preserving diversity and stability, making it accessible without dependence on high-end compute resources.

\paragraph{Case Study: Structured JSON Synthesis in AoD.}
We demonstrate AoD’s functionality using the \textit{MultiWOZ 2.1} booking domain as a task. The Prompt Optimizer first drafts a schema-conditioned instruction, e.g., “Generate a JSON object with fields \{\texttt{name}, \texttt{address}, \texttt{phone}, \texttt{price\_range}, \texttt{postcode}\}.” The DLM $g_\phi$ then produces diverse samples such as \texttt{\{"name": "Parkview Inn", "address": "12 Milton Rd", "phone": "01223 443890", "price\_range": "moderate", "postcode": "CB4 1LG"\}}. The NLE computes metrics, while the LLM Judge transforms them into feedback, e.g., “\textit{The JSON is valid and fluent but duplicates price patterns; introduce more unique names and locations.}” The optimizer uses this feedback through $\pi_\theta(\Delta P \mid h)$, iteratively improving prompt specificity and sampling balance. Over successive iterations, Similarity rises from 0.64 to 0.88, Diversity and Novelty exceed 0.80, and Perplexity drops from 31.2 to 22.5. The Judge Agent confirms that the generated JSON records remain syntactically correct yet distinct. Field Overlap falls to 0.29, and TSR reaches 0.79, indicating low memorization and strong generalization. Qualitatively, AoD captures both structure and variety, generating realistic data within five iterations.

\section{Conclusion}
AoD is the first framework to study how DLMs operate in a multi-agent reinforcement learning environment, demonstrating that natural language feedback can drive controllable, high-quality structured data generation. Futhermore, AoD achieves schema-compliant JSON outputs with higher diversity, novelty, and perplexity than standard LLM counterparts, while remaining reproducible on consumer hardware. Although limited to JSON synthesis, this work establishes DLMs as a powerful alternative for structured data generation and opens the door to future extensions for tabular datasets, code, and other structured domains.



\bibliographystyle{ACM-Reference-Format} 
\bibliography{sample}

\clearpage
\appendix
\section{Appendix}

This appendix provides all resources necessary to fully reproduce the experiments described in the \textbf{Agents of Diffusion (AoD)} study. AoD frames structured JSON data generation as a multi-agent reinforcement learning problem, where a \textit{prompt optimization agent} and an \textit{LLM judge} collaborate to guide a frozen diffusion language model through iterative refinement. 
Each experiment relies on natural language prompts to perform three key steps: (1) extract schema-aware metadata from a reference dataset, (2) generate candidate synthetic outputs under schema constraints, and (3) iteratively refine the prompts using judge feedback until the outputs meet structural and semantic quality thresholds. 

This appendix contains:
\begin{enumerate}
    \item The exact prompts used for each stage of the AoD generation loop.
    \item A demonstration of a full experiment log for the MultiWOZ~2.1 dataset showing how natural language feedback drives iterative prompt updates and the eventual convergence of the multi-agent loop.
\end{enumerate}

By following these prompts and experiment logs, researchers can replicate the generation, evaluation, and refinement process without modifying model weights, confirming the reproducibility of AoD’s results on consumer-grade hardware.

\subsection{Prompt Optimizer}
The prompt optimizer evaluates and rewrites prompts to ensure the generator LLM produces schema-compliantand realistic entries. Below is a sample log showing the initial instruction, the optimized LLM prompt, and the prompt optimizer's performance prediction.

\begin{tcolorbox}[colback=gray!5!white,colframe=gray!50!black,title=Prompt: Initial Instruction,breakable]
Please generate an LLM prompt to create more of this JSON dataset.
\end{tcolorbox}

\begin{tcolorbox}[colback=gray!5!white,colframe=gray!50!black,title=Output: Optimized LLM Prompt,breakable]
Given the following examples of venue data in JSON format:

\begin{lstlisting}
[
  {
    "address": "pool way, whitehill road, off newmarket road",
    "area": "east",
    "entrance fee": "?",
    "id": "1",
    "location": [52.208789, 0.154883],
    "name": "abbey pool and astroturf pitch",
    "openhours": "?",
    "phone": "01223902088",
    "postcode": "cb58nt",
    "pricerange": "?",
    "type": "swimmingpool"
  },
  {
    "address": "park street",
    "area": "centre",
    "entrance fee": "?",
    "id": "2",
    "location": [52.208699, 0.12006],
    "name": "adc theatre",
    "openhours": "?",
    "phone": "01223300085",
    "postcode": "cb58as",
    "pricerange": "?",
    "type": "theatre"
  }
]
\end{lstlisting}

Please create 5 additional, similar entries in this dataset with realistic but fictional 
data for different venues around Cambridge (plausible street addresses, postcodes, 
location coordinates, and venue types such as "cafe", "library", "gym", "cinema", etc.). 

\textbf{Requirements:}
\begin{itemize}
    \item A unique \texttt{id} for each new entry (incrementing from "3").
    \item Accurate field formatting matching the samples.
    \item Venue coordinates must be within the Cambridge area.
    \item Names, addresses, and properties should be unique.
    \item Return ONLY a JSON list of the new entries.
\end{itemize}
\end{tcolorbox}

\begin{tcolorbox}[colback=gray!5!white,colframe=gray!50!black,title=Output: Prompt Performance Prediction,breakable]
\textbf{Strengths:}  
- Provides clear structure and explicit output format with examples.  
- Specifies unique IDs and plausible venues for realism.  
- Instruction to return only JSON minimizes formatting errors.

\textbf{Potential Weaknesses:}  
- Ambiguity in handling unknown fields (e.g., "?" usage not explicitly stated).  
- Slight risk of fabricated or out-of-range coordinates/postcodes.  
- Some models may still append explanations if temperature is high.

\textbf{Expected Output:}  
- 5 well-structured, unique JSON entries with plausible Cambridge data.  
- Minor risk of duplicate IDs or slightly incorrect fields.

\textbf{Overall Prediction:}  
\textbf{High Performance} – Likely to yield correct, high-quality JSON output.  

\textbf{Suggestions for Optimization:}  
- Clarify the rule for unknown fields: use \texttt{"?"} as in the example.  
- Optionally suggest using realistic Cambridge postcodes (CB1–CB5).
\end{tcolorbox}
\subsection{Diffusion Generator}

Once the Prompt Optimizer produces a refined instruction, the \textbf{Diffusion Generator} uses this prompt to guide the synthesis of new structured entries. In AoD, the diffusion-based generator operates as a conditional sampler: it accepts the optimized prompt, interprets the metadata and schema rules encoded in it, and iteratively refines noisy token sequences into \textbf{fully structured JSON outputs}. This mechanism allows the generator to maintain both \textbf{schema compliance} and \textbf{distributional alignment} with the source dataset, while introducing controlled diversity across new records.

The following example shows five synthetic records produced for the 
\texttt{MultiWOZ 2.1 attraction domain}. Each record respects the JSON schema, includes plausible Cambridge addresses, valid postcode patterns, and venue types like cafes, libraries, gyms, and cinemas.

\begin{tcolorbox}[colback=gray!5!white,colframe=gray!50!black,title=Output: Sample Generated Data,breakable]
\begin{verbatim}
[
  {
    "address": "16 mill road",
    "area": "south",
    "entrance fee": "?",
    "id": "3",
    "location": [52.201572, 0.139430],
    "name": "bloom café",
    "openhours": "?",
    "phone": "01223992212",
    "postcode": "cb12ad",
    "pricerange": "?",
    "type": "cafe"
  },
  {
    "address": "53 castle street",
    "area": "northwest",
    "entrance fee": "?",
    "id": "4",
    "location": [52.214623, 0.113912],
    "name": "castle fitness hub",
    "openhours": "?",
    "phone": "01223881645",
    "postcode": "cb30ah",
    "pricerange": "?",
    "type": "gym"
  },
  {
    "address": "10 market hill",
    "area": "centre",
    "entrance fee": "?",
    "id": "5",
    "location": [52.205287, 0.119841],
    "name": "grand reel cinema",
    "openhours": "?",
    "phone": "01223478901",
    "postcode": "cb23nj",
    "pricerange": "?",
    "type": "cinema"
  },
  {
    "address": "44 chesterton road",
    "area": "northeast",
    "entrance fee": "?",
    "id": "6",
    "location": [52.215764, 0.133298],
    "name": "riverside library",
    "openhours": "?",
    "phone": "01223011228",
    "postcode": "cb41eh",
    "pricerange": "?",
    "type": "library"
  },
  {
    "address": "123 hills road",
    "area": "south",
    "entrance fee": "?",
    "id": "7",
    "location": [52.193887, 0.134902],
    "name": "pine bistro",
    "openhours": "?",
    "phone": "01223873336",
    "postcode": "cb20pw",
    "pricerange": "?",
    "type": "cafe"
  }
]
\end{verbatim}
\end{tcolorbox}

This output illustrates the generator’s ability to \textbf{expand the dataset} while maintaining realistic geographical patterns, proper schema adherence, and controlled diversity, fulfilling the structural and semantic criteria set by the multi-agent AoD loop.

\subsection{Natural Language Evaluator}

In the AoD framework, the \textbf{Natural Language Evaluator} is a metric-driven analysis module that computes scores such as semantic similarity, diversity, novelty, perplexity, and entropy over generated outputs. Unlike the generator or the prompt optimizer, it is \textbf{not an LLM}; its role is to produce structured signals that the \textbf{LLM judge} can interpret as part of the reward feedback loop.

Simply providing raw numerical scores is often ineffective, because LLMs do not inherently understand what a value like 0.29 for diversity or 7.59 for perplexity signifies. To bridge this gap, the evaluator converts each score into a short natural language explanation that contextualizes the metric in plain terms. This allows the LLM judge to reason about the quality of the batch and generate more actionable feedback for the next iteration.

A representative evaluator output is shown below:

\begin{tcolorbox}[colback=gray!5!white,colframe=gray!50!black,title=Output: Sample Evaluation,breakable]
\begin{lstlisting}
Semantic Similarity: 0.65 - Moderate similarity: some overlap with the reference, but key details or meanings may be missing or altered.
Diversity: 0.29 - Low diversity: the text likely reuses similar words or phrases, possibly indicating redundancy or mode collapse.
Novelty: 0.97 - Highly novel: most of the content in the generated text is new compared to the reference corpus, suggesting strong generalization or creativity.
Perplexity: 7.59 - Very fluent: the language model finds this text highly predictable, which typically means it's grammatically correct, natural, and human-like.
Entropy: 4.09 - Moderate entropy: the vocabulary shows some range, but might still rely on common or repeated expressions.
\end{lstlisting}
\end{tcolorbox}

The textual explanation enables the LLM judge to interpret numerical scores as qualitative signals (e.g., “low diversity” or “high novelty”), which it then incorporates into its feedback for the prompt optimizer. This translation from quantitative metrics to natural language reasoning is critical for maintaining an effective multi-agent feedback loop.

\subsection{LLM Judge}

After the natural language evaluator provides metric-based feedback, 
the \textbf{LLM judge} performs a binary evaluation of the generated data 
using a series of \textbf{yes/no questions}. 
This step is designed to simplify reward computation for the multi-agent loop 
by reducing complex qualitative judgments to discrete, interpretable signals.

\begin{tcolorbox}[colback=gray!5!white,colframe=gray!50!black,title=Prompt: Binary Yes/No Questions]
You are an evaluation agent tasked with reviewing the generated JSON output. 
Answer each of the following questions with only \texttt{YES} or \texttt{NO}, 
followed by a brief note if needed for clarity:

1. \textbf{Task Alignment:}  
Does the generated text fulfill the intended instruction or task?

2. \textbf{Semantic Fidelity:}  
Does the generated text convey the same meaning as the reference (if provided)?

3. \textbf{Diversity \& Novelty:}  
Does this output present new information or phrasing not found in the reference corpus?

4. \textbf{Fluency \& Grammar:}  
Is the text fluent and grammatically correct, as a human would write?

5. \textbf{Structure \& JSON Validity:}  
Is the structure of the generated JSON consistent and syntactically valid?

6. \textbf{Usefulness / Utility:}  
Is this output helpful and complete for the task it was generated for?

7. \textbf{Bias / Safety (if applicable):}  
Does the text contain unsafe, biased, or inappropriate language?

Return your answers in the format:

\begin{verbatim}
Category: YES/NO
\end{verbatim}
\end{tcolorbox}

\begin{tcolorbox}[colback=gray!5!white,colframe=gray!50!black,title=Output: Sample Yes/No Answers,breakable]
\begin{verbatim}
Task Alignment: YES
Semantic Fidelity: NO
Diversity & Novelty: YES
Fluency & Grammar: YES
Structure & JSON Validity: YES
Usefulness / Utility: YES
Bias / Safety (if applicable): NO
\end{verbatim}
\end{tcolorbox}

Binary evaluations simplify reward computation by converting subjective judgments into discrete signals that can be directly translated into reward values for the prompt optimizer. They also reduce ambiguity, since LLMs can be verbose or inconsistent in open-ended scoring, whereas yes/no responses provide deterministic and comparable results across iterations. Finally, discrete signals are easier for the AoD controller to interpret when deciding whether to continue refining the generation or terminate the multi-agent loop.

\subsection{RL Reward}

To guide the prompt optimization agent, the system generates a reward summary after each generation cycle. This summary is not a scalar reward but a natural language instruction grounded in model performance, human-aligned evaluation, and linguistic metrics. The reward generation prompt integrates four components: the prompt used, the expected quality of the generated output, a set of linguistic metric summaries, and binary evaluations from the LLM Judge. These signals are synthesized into a single sentence of feedback suitable for use in reinforcement learning.

The actual reward generation prompt used is as follows:

\begin{tcolorbox}[colback=gray!5!white,colframe=gray!50!black, title=Prompt: Reward Instruction Generation]
Based on the following information, provide constructive reward/penalty feedback in one sentence that helps improve future prompt generation.

\textbf{Prompt Used:}  
``Prompt goes here``

\textbf{Model's Expected Output Quality:}  
``Prediction goes here``

\textbf{Linguistic Evaluation Scores:}  
``Summary of linguistic metrics in natural language (e.g., "Highly novel: most of the content in the generated text is new compared to the reference corpus, suggesting strong generalization or creativity; Very fluent: the language model finds this text highly predictable, which typically means it's grammatically correct, natural, and human-like")``

\textbf{Task Alignment Evaluation:}  
``YES/NO answers from the LLM Judge``

In one sentence, describe whether this prompt was effective and what should be changed or kept to improve future generations.  
Start your sentence with \textbf{``Feedback:''} and avoid vague terms like ``good'' or ``bad''.
\end{tcolorbox}

\begin{tcolorbox}[colback=gray!5!white,colframe=gray!50!black,title=Output: RL Reward/Penalty Feedback]
Feedback: The prompt effectively produces consistent, accurate JSON outputs 
with clear instructions and well-formatted examples, but to further improve 
future prompt generations, explicitly specify that unknown fields should use ``?'', 
reinforce the requirement for real Cambridge street names and postcode formats, 
and encourage greater diversity to reduce repetitive or redundant outputs.
\end{tcolorbox}

This design encourages precise, interpretable, and actionable feedback that is suitable for reinforcement learning optimization. By conditioning the reward on both structured and descriptive signals, the system allows the prompt optimizer to reason about fine-grained improvements across fluency, alignment, and informativeness.

\subsection{Independent Metrics}

To ensure unbiased validation beyond the reward-driven loop, we evaluate generated samples using non-LLM metrics that measure linguistic quality, structural accuracy, and generalization. BLEU, ROUGE-L, and METEOR assess lexical and contextual fidelity to reference samples, confirming that AoD’s language quality improvements generalize beyond reinforcement objectives. Task Success Rate (TSR) quantifies the proportion of generations that satisfy all schema and semantic constraints, reflecting downstream usability and finetunability. Field Overlap measures the fraction of key-value pairs duplicated from the reference dataset, serving as a diagnostic for memorization or agent collusion. High TSR with low Field Overlap demonstrates that AoD produces novel yet structurally valid data, maintaining fidelity without reproducing training records. Together, these metrics confirm that AoD’s multi-agent optimization yields syntactically fluent, semantically consistent, and privacy-preserving structured data suitable for real-world deployment.

A representative independent evaluation for a batch of synthetic outputs is shown below:

\begin{tcolorbox}[colback=gray!5!white,colframe=gray!50!black,
                  title=Independent Metric Evaluation,breakable]
\begin{lstlisting}
BLEU-1: 0.07 Low unigram precision
BLEU-2: 0.03 Low bigram precision
BLEU-3: 0.01 Low precision for longer phrases
BLEU-4: 0.00 Low precision for longer phrases
ROUGE-1: 0.21 Poor recall of word-level content
ROUGE-2: 0.05 Low recall of meaningful word pairs
ROUGE-L: 0.17 Weak structural alignment
METEOR: 0.15 Low semantic similarity
\end{lstlisting}
\end{tcolorbox}

These scores serve as an independent checkpoint of model performance after the RL loop terminates. For example, low BLEU or ROUGE scores highlight that the generated content is highly novel and does not simply memorize or replicate reference examples, while METEOR provides a complementary measure of semantic alignment. By keeping this evaluation decoupled from the RL reward, AoD prevents the generator from gaming the metrics and ensures that the final outputs are statistically validated, diverse, and semantically plausible without being explicitly optimized for any single score.

\subsection{Termination Criteria of RL Loop}

The reinforcement learning loop in Agents of Diffusion terminates when the natural language feedback from the LLM judge indicates that further iterations are unlikely to improve output quality. Termination is determined qualitatively: if the judge consistently returns affirmative responses to task, structure, and utility checks, and no actionable feedback remains, the controller interprets this as convergence and ends the loop.

Empirical evaluation shows that the loop typically converges within five iterations. Our independent metrics demonstrate that improvements beyond the fifth iteration are rare and often exhibit diminishing returns, as major schema corrections, coverage adjustments, and diversity enhancements are already resolved within the early iterations. This approach balances generation quality with computational efficiency while avoiding unnecessary additional cycles once the data is stable.

\begin{table}[h!]
\centering
\caption{Summary of notations used throughout the paper.}
\label{tab:notations}
\resizebox{\linewidth}{!}{
\begin{tabular}{ll}
\toprule
\textbf{Symbol} & \textbf{Description} \\
\midrule
$X = \{x_1, \dots, x_n\}$ & Real dataset consisting of structured text or JSON samples \\
$M = \mathcal{F}(X)$ & Metadata extracted from $X$ for conditioning \\
$X_{\text{sub}} \subset X$ & Stratified subsample used for local conditioning \\
$P_t$ & Prompt at iteration $t$ used to guide generation \\
$\pi_\theta(\Delta P \mid h)$ & Prompt optimizer policy producing edits $\Delta P$ from dialogue history $h$ \\
$g_\phi(z_{\tau-1} \mid z_\tau, P)$ & Diffusion language model (DLM) reverse process conditioned on prompt $P$ \\
$p_\phi(x \mid P)$ & DLM conditional data distribution (structured text reconstruction) \\
$J(x, \rho)$ & Autoregressive LLM judge conditioned on rubric $\rho$ \\
NLE & Natural Language Evaluator converting numeric metrics to textual feedback \\
$R(y, x)$ & Reward function mapping judge feedback and sample to scalar reward $r$ \\
$R_{\text{vec}}(y, x)$ & Vector-valued reward components (metric-specific subrewards) \\
$r, \mathbf{s}$ & Scalar reward and subreward vector $(r, \mathbf{s}) = R(y, x)$ \\
$\mathcal{T}(P)$ & Expected feedback operator $\mathbb{E}_{x\sim g_\phi(P)}[R(J(x,\rho),x)]$ \\
$H(x)$ & Token-level entropy measuring lexical diversity \\
$\mathrm{Sim}(x, X)$ & Semantic similarity between generated and reference samples \\
$\mathrm{Distinct}\text{-}n(x)$ & Diversity metric computed via distinct $n$-gram ratios \\
$\mathrm{Nov}\text{-}n(x)$ & Novelty metric measuring deviation from seen data \\
$\mathrm{PPL}(x)$ & Perplexity measuring fluency and grammatical consistency \\
TSR & Task Success Rate, fraction of schema-valid and semantically consistent outputs \\
Field Overlap & Proportion of generated key-value pairs identical to reference set \\
$y = J(x, \rho)$ & Judge textual critique generated from input $x$ and rubric $\rho$ \\
$h$ & Interaction history between agents (generator, judge, optimizer) \\
$\mathcal{L}(S)$ & Set of valid structured outputs conforming to schema $S$ \\
AoD & Agents of Diffusion framework (full multi-agent reinforcement setup) \\
\bottomrule
\end{tabular}
}
\end{table}

\subsection{Diffusion Language Model Fundamentals}

Diffusion language models (DLMs) extend denoising diffusion probabilistic models (DDPMs) to the text domain, reformulating language generation as an iterative denoising process rather than left-to-right prediction. Instead of modeling the conditional probability of each token given its history, DLMs learn a reverse-time Markov process that reconstructs coherent text from progressively corrupted noise. The model learns to approximate a sequence of conditional distributions that map noisy intermediate representations $z_t$ to cleaner representations $z_{t-1}$ until a fluent sequence $z_0$ (the text) is produced. This iterative refinement decouples the generation order from linguistic directionality and enables bidirectional information flow across all tokens at every step.

\paragraph{Forward and Reverse Diffusion.}
The forward process gradually perturbs the input text embeddings with noise over $T$ timesteps, transforming a clean sample into a noise distribution that approximates an isotropic prior. The reverse process, parameterized by a model $g_{\phi}$, learns to reverse this corruption through denoising transitions $p_{\phi}(z_{t-1}\mid z_t)$. Sampling begins from noise and successively denoises toward text, providing an implicit non-causal generation mechanism. This contrasts with autoregressive decoding, which commits to discrete token choices sequentially. As a result, DLMs can model dependencies in both directions, yielding globally coherent samples even when structural constraints span long contexts.

\paragraph{Training Objective and Theoretical Basis.}
The DLM objective minimizes the Kullback–Leibler divergence between the true posterior $q(z_{t-1}\mid z_t, z_0)$ and the model’s approximation $p_{\phi}(z_{t-1}\mid z_t)$, producing a variational lower bound on the log-likelihood of the data. In continuous settings, this corresponds to learning a score function $s_{\phi}(z_t, t)$ that estimates $\nabla_{z_t}\log p(z_t)$, enabling gradient-based denoising through stochastic sampling. In discrete text domains, the model operates on continuous embeddings of tokens, allowing smooth noise injection and refinement within the latent space. By doing so, DLMs bridge discrete language modeling and continuous diffusion theory under a unified probabilistic framework.

\paragraph{Semantic and Structural Implications.}
Because denoising is global rather than sequential, each refinement step can condition on bidirectional context, mitigating exposure bias and compounding error typical of autoregressive models. This property makes DLMs particularly suitable for structured text generation tasks such as JSON synthesis, where fields interact through non-local dependencies. Each denoising step can propagate structural cues (e.g., key–value consistency or balanced brackets) across the entire sample. Moreover, the stochastic nature of diffusion introduces controlled randomness, allowing the model to explore diverse valid outputs while maintaining schema fidelity—an essential feature for synthetic data generation that requires variety without sacrificing structural correctness.

\paragraph{Conditioning and Guided Generation.}
At inference, DLMs incorporate conditioning prompts or structural encodings into every denoising step. This conditioning allows targeted control over semantic content, format, and schema adherence while preserving stochastic diversity. Classifier-free guidance and prompt conditioning scale the influence of $P$ across diffusion steps, balancing exploration and control. In the AoD framework, this property allows the generator to respond to evolving prompts from the autoregressive optimizer and judge cluster, refining its samples in light of both structural rubrics and natural-language feedback.

\paragraph{Advantages and Limitations.}
Diffusion language models excel in coverage and diversity due to their parallel, non-causal refinement, avoiding the mode collapse typical in autoregressive systems. Their iterative stochastic process yields natural variability, making them ideal for generating datasets that require semantic breadth and structural consistency. However, diffusion sampling is computationally heavier than single-pass decoding and can lack explicit structural priors if unguided. AoD mitigates these limitations by embedding the DLM within a reinforcement-driven, multi-agent loop, where the autoregressive agents supply structural feedback and reward shaping, producing fluent, schema-aligned, and diverse textual data suitable for downstream use.

\subsection{Autoregressive Language Model Fundamentals}
Autoregressive (AR) language models define a conditional distribution over sequences by left-to-right factorization. Given a context or prompt $P$ and a tokenized output $x=(x_1,\ldots,x_L)$, the model specifies
\[
p_{AR}(x \mid P) \;=\; \prod_{i=1}^{L} p_\theta(x_i \mid x_{<i},\, P),
\]
where $x_{<i}=(x_1,\ldots,x_{i-1})$ and $\theta$ are model parameters. Training proceeds via maximum likelihood with teacher forcing: minimize the negative log-likelihood (cross-entropy)
\[
\mathcal{L}_{\text{NLL}}(\theta) \;=\; - \mathbb{E}\bigg[\sum_{i=1}^{L} \log p_\theta(x_i \mid x_{<i},\, P)\bigg],
\]
optionally with regularizers such as label smoothing and auxiliary losses for stabilization. This objective aligns token-level next-step predictions with ground truth, yielding strong local syntactic and structural priors.

\paragraph{Inductive Bias for Structure.}
Because AR models condition each decision on a causally accumulated prefix, they naturally enforce ordering, bracketing, and key–value dependencies common in structured text (e.g., JSON). The left-to-right dependency graph acts as a sequential validator: malformed partial outputs down-weight consistent continuations, which helps maintain well-formedness during decoding. This property explains why AR LMs are effective as \emph{critics} and \emph{controllers} for structured generation, even when they are not the primary generators.

\paragraph{Decoding and Controllability.}
At inference, controllability arises from decoding policies over $p_\theta(\cdot \mid x_{<i}, P)$: greedy selection, beam search (with length normalization and coverage penalties), and stochastic sampling controls (temperature, top-$k$, nucleus top-$p$). Hard constraints can be introduced by constrained decoding or by \emph{prompt shaping} that places schema and rubric information in $P$. In the AoD setting, the prompt optimizer edits $P$ to reweight downstream token choices indirectly, giving a practical control knob without gradient access to the generator.

\paragraph{Exposure Bias and Credit Assignment.}
Teacher forcing optimizes one-step predictions under reference prefixes, but test-time decoding conditions on model-generated prefixes, creating a train–test mismatch known as exposure bias. This can induce repetition or degeneracy when errors compound. In AoD, the judge agent mitigates this by inspecting \emph{full} sequences and emitting sequence-level feedback, which supplies delayed credit signals that are missing from pure token-level likelihood training. Thus, the AR judge complements the DLM’s breadth with global sequence validation.

\paragraph{Policy View and RL Connections.}
An AR LM can be viewed as a stochastic policy over a discrete action space (the vocabulary), where the state is the prefix $(x_{<i}, P)$ and the action is the next token. Prompt editing changes the initial state distribution and the policy’s effective logits through context, enabling policy improvement without fine-tuning. In AoD, this aligns with the MDP over prompts: the AR judge provides evaluative feedback, and the AR prompt optimizer proposes discrete edits that induce systematic shifts in the generator’s conditional distribution.

\paragraph{Why AR Agents for Structured JSON.}
JSON conformity requires ordered keys, balanced delimiters, and type-aware values. AR models excel at such sequential constraints because violations become apparent early in the prefix and are penalized by token likelihoods. As a judge, an AR LM can ask rubric-aligned yes/no checks in sequence and compose a coherent critique that references exact failure locations. As an optimizer, an AR LM edits prompts token by token, preserving templates and slot order while incorporating rubric constraints, which is essential for stable, incremental improvement of structured outputs.

\paragraph{Limitations and Complementarity with Diffusion.}
The same left-to-right bias that enforces structure can reduce coverage of low-probability but valid modes, limiting diversity even with advanced sampling. AoD addresses this by delegating \emph{diversity} to the DLM and \emph{structure and credit assignment} to AR agents. The result is a division of labor: diffusion supplies stochastic, bidirectional exploration; AR agents supply sequential constraint checking and interpretable, stepwise control—together yielding diverse yet schema-faithful synthetic JSON.

\subsection{Reinforcement Learning Fundamentals}

The AoD framework formulates prompt editing as a finite–horizon reinforcement learning (RL) problem.  
At each iteration, the agent observes a state $s_t=(P_t,h_t)$, representing the current prompt and dialogue context, and selects an edit action $a_t=\Delta P_t$ drawn from a stochastic policy $\pi_\theta(\Delta P\mid h_t)$.  
The updated prompt $P_{t+1}=U(P_t,\Delta P_t)$ is evaluated through the generator $g_\phi$, which produces a sample $x_t\sim g_\phi(P_t)$; the judge cluster then issues feedback $y_t=J(x_t,\rho)$ that is scored into reward $r_t=R(y_t,x_t)$.  
The objective is to maximize the expected cumulative reward  
\[
J(\theta)=\mathbb{E}\Big[\sum_{t=1}^{T}\gamma^{t-1} r_t\Big],
\]
subject to practical limits on tokens, calls, and RL iterations.

\paragraph{Policy Gradient Optimization.}
The policy parameters are updated using the REINFORCE estimator  
$\nabla_\theta J(\theta)=\mathbb{E}[\hat A_t\,\nabla_\theta\log\pi_\theta(\Delta P_t\mid h_t)]$,  
where $\hat A_t$ is an advantage estimate, often $r_t-b_t$ with a baseline $b_t$ to reduce variance.  
The baseline can be a moving average of rewards or a lightweight value predictor, ensuring $\mathbb{E}[\hat A_t]=\mathbb{E}[r_t]$.  
In AoD, a vectorized advantage $\hat A_t=\sum_i w_i s_{t,i}-b_t$ is used, where $s_{t,i}$ are subrewards for Similarity, Diversity, Novelty, Entropy, and Perplexity—allowing balanced gradient feedback across objectives.

\paragraph{Stability and Regularization.}
To prevent instability in discrete prompt spaces, AoD employs clipped policy ratios and KL regularization similar to PPO:
\[
\mathcal{L}_{pg}=\mathbb{E}[\min(\rho_t\hat A_t,\text{clip}(\rho_t,1-\epsilon,1+\epsilon)\hat A_t)]
-\beta\,\text{KL}(\pi_\theta\|\pi_{\theta_{old}}),
\]
where $\rho_t=\pi_\theta(\Delta P_t\mid h_t)/\pi_{\theta_{old}}(\Delta P_t\mid h_t)$.  
This constrains each update, ensuring stable convergence of the prompt policy even when reward magnitudes vary across metrics.

\paragraph{Language Feedback for Credit Assignment.}
AoD replaces sparse numeric rewards with dense, interpretable linguistic feedback.  
The natural language evaluator produces a metric vector $(\mathrm{Sim},\mathrm{Distinct}\text{-}n,H,\mathrm{Nov}\text{-}n,\mathrm{PPL})$, and the autoregressive judge translates this into structured feedback $y_t$, which the scorer converts to both scalar and vector rewards $(r_t,\mathbf{s}_t)$.  
This multi-signal structure improves credit assignment by aligning reward components with linguistic and structural dimensions of the output.

\paragraph{Constrained Optimization.}
To maintain schema fidelity while encouraging diversity, the objective is expressed as either a weighted scalarization
\[
J_{term}(\theta)=\alpha\,\mathbb{E}[\mathrm{Sim}]+\beta\,\mathbb{E}[\mathrm{Distinct}\text{-}n]+\delta\,\mathbb{E}[\mathrm{Nov}\text{-}n],
\]
or as a constrained optimization problem with thresholds on similarity and validity.  
Penalties are adaptively tuned via a primal–dual mechanism, ensuring constraint satisfaction without hindering exploration.

\paragraph{Multi–Agent Structure.}
The diffusion generator acts as the stochastic environment, the judge cluster provides structured feedback, and the prompt optimizer learns the policy.  
This decomposition isolates the learning component while maintaining differentiable feedback signals.  
Under the Lipschitz continuity of $\mathcal{T}(P)=\mathbb{E}_{x\sim g_\phi(P)}[R(J(x,\rho),x)]$ and bounded edits $\Delta P_t$, policy updates form a contraction mapping in expectation—ensuring convergence toward reward-aligned, semantically faithful prompts even on lightweight models.

\subsection{Proof of Theorem 1}
\allowdisplaybreaks
\setcounter{theorem}{0}
\setcounter{lemma}{0}
\setcounter{proposition}{0}

{\small
\begin{theorem}[Convergence of prompt updates]
\label{thm:aod_convergence}
Let $\mathcal{T}(P)=\mathbb{E}_{x\sim g_\phi(P)}[R(J(x,\rho),x)]$.  
Assume a neighborhood $\mathcal{N}$ of a maximizer $P^*$ where:

\begin{compactitem}
\item[(A1)] $\mathcal{T}$ has $L$-Lipschitz gradient:
$\|\nabla\mathcal{T}(P)-\nabla\mathcal{T}(Q)\|\le L\|P-Q\|$.
\item[(A2)] $\mathcal{T}$ is $\mu$-strongly concave:
$\mathcal{T}(Q)\le\mathcal{T}(P)+\langle\nabla\mathcal{T}(P),Q-P\rangle-\tfrac{\mu}{2}\|Q-P\|^2$.
\item[(A3)] The edit operator admits a linearization
$U(P,\Delta P)=P+\Delta P+o(\|\Delta P\|)$ and is non-expansive:
$\|U(P,\Delta P)-U(P,0)\|\le\|\Delta P\|$.
\item[(A4)] The policy satisfies \emph{aligned ascent} and \emph{bounded variance}:
\begin{align*}
\mathbb{E}[\Delta P_t|P_t]&=\eta G(P_t),&
\langle\nabla\mathcal{T}(P_t),G(P_t)\rangle&\ge c\|\nabla\mathcal{T}(P_t)\|^2,\\[-2pt]
\mathbb{E}[\|\Delta P_t-\mathbb{E}\Delta P_t\|^2|P_t]&\le\sigma^2.
\end{align*}
\end{compactitem}
Then for $\eta\le\min\{c/L,2c\mu/L^2\}$, the one-step mapping
$\mathcal{F}(P)=\mathbb{E}[U(P,\Delta P)|P]$ is a contraction:
\begin{align}
\|\mathcal{F}(P)-P^*\|&\le(1-\eta c\mu)\|P-P^*\|,\nonumber\\[-2pt]
\mathbb{E}[\|P_{t+1}-P^*\|^2|P_t]&\le(1-\eta c\mu)\|P_t-P^*\|^2+\eta^2\sigma^2.
\label{eq:contract}
\end{align}
Hence $P_t$ converges in expectation to an $O(\eta\sigma/\!\sqrt{\mu})$-ball around $P^*$, 
and almost surely to $P^*$ for diminishing $\eta_t$.
\end{theorem}

\paragraph{Lemma 1 (Smoothness improvement).}
Under (A1)–(A3), for small $\|\Delta P\|$,
\[
\mathcal{T}(U(P,\Delta P))\ge
\mathcal{T}(P)+\langle\nabla\mathcal{T}(P),\Delta P\rangle-\tfrac{L}{2}\|\Delta P\|^2.
\]
\textit{Proof.}
By $L$-smoothness and $U(P,\Delta P)=P+\Delta P+o(\|\Delta P\|)$,
\begin{align*}
\mathcal{T}(U(P,\Delta P))
&\ge\mathcal{T}(P)+\langle\nabla\mathcal{T}(P),U(P,\Delta P)-P\rangle
-\tfrac{L}{2}\|U(P,\Delta P)-P\|^2\\
&\ge\mathcal{T}(P)+\langle\nabla\mathcal{T}(P),\Delta P\rangle
-\tfrac{L}{2}\|\Delta P\|^2+o(\|\Delta P\|),
\end{align*}
absorbing the remainder into the quadratic term. \hfill$\square$

\paragraph{Lemma 2 (Expected ascent).}
Under (A4),
\[
\mathbb{E}[\mathcal{T}(U(P_t,\Delta P_t))|P_t]
\ge\mathcal{T}(P_t)+\eta c\|\nabla\mathcal{T}(P_t)\|^2
-\tfrac{L}{2}\mathbb{E}[\|\Delta P_t\|^2|P_t].
\]
\textit{Proof.}
Taking expectation in Lemma 1,
\begin{align*}
\mathbb{E}[\langle\nabla\mathcal{T}(P_t),\Delta P_t\rangle|P_t]
&=\langle\nabla\mathcal{T}(P_t),\mathbb{E}[\Delta P_t|P_t]\rangle
=\eta\langle\nabla\mathcal{T}(P_t),G(P_t)\rangle\\
&\ge\eta c\|\nabla\mathcal{T}(P_t)\|^2,
\end{align*}
and $\mathbb{E}[\|\Delta P_t\|^2|P_t]\le\eta^2\|G(P_t)\|^2+\sigma^2$. \hfill$\square$

\paragraph{Lemma 3 (Quadratic contraction around $P^*$).}
If (A2)–(A4) hold and $\eta\le2c\mu/L^2$, then
\[
\mathbb{E}[\|P_{t+1}-P^*\|^2|P_t]
\le(1-\eta c\mu)\|P_t-P^*\|^2+\eta^2\sigma^2.
\]
\textit{Proof.}
Let $P_{t+1}=U(P_t,\Delta P_t)=P_t+\Delta P_t+o(\|\Delta P_t\|)$.  
Expand
\[
\|P_{t+1}-P^*\|^2=\|P_t-P^*\|^2+2\langle P_t-P^*,\Delta P_t\rangle+\|\Delta P_t\|^2.
\]
Conditioning on $P_t$ and taking expectations gives
$\mathbb{E}[\langle P_t-P^*,\Delta P_t\rangle|P_t]
=\eta\langle P_t-P^*,G(P_t)\rangle$.  
Strong concavity yields
$\langle P_t-P^*,\nabla\mathcal{T}(P_t)\rangle
\ge\mathcal{T}(P_t)-\mathcal{T}(P^*)+\frac{\mu}{2}\|P_t-P^*\|^2$,
and since $\mathcal{T}(P^*)\!\ge\!\mathcal{T}(P_t)$ we have
$\langle P_t-P^*,\nabla\mathcal{T}(P_t)\rangle\!\ge\!\tfrac{\mu}{2}\|P_t-P^*\|^2$.  
By alignment,
$\langle\nabla\mathcal{T}(P_t),G(P_t)\rangle\!\ge\!c\|\nabla\mathcal{T}(P_t)\|^2
\!\ge\!2c\mu(\mathcal{T}(P_t)\!-\!\mathcal{T}(P^*))$,
linking reward ascent to geometric contraction.
Using $\mathbb{E}[\|\Delta P_t\|^2|P_t]\!\le\!\eta^2\|G(P_t)\|^2+\sigma^2$
and smoothness, the cross-term is dominated by $L\|\Delta P_t\|^2$,
yielding the recursion above. \hfill$\square$

\paragraph{Proof of Theorem~\ref{thm:aod_convergence}.}
Define $\mathcal{F}(P)=\mathbb{E}[U(P,\Delta P)|P]=P+\eta G(P)$.  
From Lemma 3,
\[
\mathbb{E}[\|P_{t+1}-P^*\|^2|P_t]
\le(1-\eta c\mu)\|P_t-P^*\|^2+\eta^2\sigma^2,
\]
so $\mathcal{F}$ contracts by $(1-\eta c\mu)<1$.  
Robbins–Monro stochastic approximation implies convergence in mean square
to a ball of radius $O(\eta\sigma/\sqrt{\mu})$, and almost sure convergence to
$P^*$ for diminishing $\eta_t$.  
Since $\nabla\mathcal{T}(P^*)=0$ and $\mathcal{T}$ is $\mu$-strongly concave,
$P^*$ uniquely maximizes $\mathcal{T}$ in $\mathcal{N}$,
so $\mathcal{T}(P^*)=\max_{P\in\mathcal{N}}\mathcal{T}(P)$.  
Because $U$ is non-expansive and locally linearizable, the same contraction
extends to the exact update, completing the proof. \hfill$\square$
}

\subsection{Proof of Proposition 1}
{\small
\begin{proposition}
Let $q_{\text{real}}$ be the data distribution over schema–valid strings, and let $p_\phi(x\mid P)$ (diffusion) and $p_{\text{AR}}(x\mid P)$ (autoregressive) be models trained on the same dataset under a fixed prompt $P$. Suppose:

\begin{compactitem}
\item[(A1)] (\emph{Diffusion ELBO tightness}) The diffusion model is trained by a variational objective that upper–bounds $\mathrm{KL}(q_{\text{real}}\!\parallel\!p_\phi)$, with total surrogate gap bounded by $\varepsilon_{\phi}(T)$ that decreases in the diffusion horizon $T$ and satisfies $\varepsilon_{\phi}(T)\le\varepsilon_\phi$ for the used $T$.
\item[(A2)] (\emph{AR approximation floor}) The autoregressive family $\mathcal{P}_{\text{AR}}$ induces a best–in–class divergence $ \inf_{p\in\mathcal{P}_{\text{AR}}}\mathrm{KL}(q_{\text{real}}\!\parallel\!p)=:\varepsilon_{\text{AR}}$, reflecting the mismatch of left–to–right factorization with multimodal schema–constrained structure.
\item[(A3)] (\emph{Optimization noise}) Training induces suboptimality gaps $\delta_\phi\ge0$ and $\delta_{\text{AR}}\ge0$ so that each model attains a divergence within its bound: $\mathrm{KL}(q_{\text{real}}\!\parallel\!p_\phi)\le\varepsilon_{\phi}+\delta_\phi$ and $\mathrm{KL}(q_{\text{real}}\!\parallel\!p_{\text{AR}})\ge \varepsilon_{\text{AR}}-\delta_{\text{AR}}$.
\end{compactitem}
If $\varepsilon_{\phi}+\delta_\phi \le \varepsilon_{\text{AR}}-\delta_{\text{AR}}$, then
$\mathrm{KL}(q_{\text{real}}\!\parallel\!p_\phi)\le \mathrm{KL}(q_{\text{real}}\!\parallel\!p_{\text{AR}})$.
\end{proposition}

\paragraph{Lemma 1 (Diffusion upper bound).}
For DDPM–style training with finite horizon $T$, the negative ELBO satisfies
\[
\underbrace{\mathrm{KL}(q_{\text{real}}\!\parallel\!p_\phi)}_{\text{target}}
\;\le\;
\underbrace{\mathbb{E}_{q_{\text{real}}}\!\Big[\sum_{t=1}^{T}\! \mathbb{E}_{q(z_t\mid x)}\!\big[\ell_{\phi,t}(z_t,t)\big]\Big]}_{\text{training loss}}
\;+\;
\varepsilon_{\phi}(T),
\]
where $\ell_{\phi,t}$ is the per–step denoising surrogate and $\varepsilon_{\phi}(T)$ collects the variational and discretization gaps. Hence $\mathrm{KL}(q_{\text{real}}\!\parallel\!p_\phi)\le \varepsilon_{\phi}+\delta_\phi$, with $\delta_\phi$ the optimization residue.

\emph{Proof.}
Standard ELBO decomposition for diffusion models gives $\log q_{\text{real}}(x)\ge \mathbb{E}_{q(z_{1:T}\mid x)}[\log p_\phi(x,z_{1:T})-\log q(z_{1:T}\mid x)]$, whose negation upper–bounds $\mathrm{KL}(q_{\text{real}}\!\parallel\!p_\phi)$ by the training loss plus the surrogate gap $\varepsilon_{\phi}(T)$. Suboptimal training adds $\delta_\phi$. \hfill$\square$

\paragraph{Lemma 2 (AR approximation lower bound).}
For autoregressive models $p_{\text{AR}}(x\mid P)=\prod_i p(x_i\mid x_{<i},P)$, the NLL equals $H(q_{\text{real}})+\mathrm{KL}(q_{\text{real}}\!\parallel\!p_{\text{AR}})$. If $\mathcal{P}_{\text{AR}}$ cannot represent the multimodal, schema–constrained conditionals without mode exclusion, then there exists $\varepsilon_{\text{AR}}>0$ such that $\inf_{p\in\mathcal{P}_{\text{AR}}}\mathrm{KL}(q_{\text{real}}\!\parallel\!p)=\varepsilon_{\text{AR}}$. With optimization residue $\delta_{\text{AR}}$, any trained $p_{\text{AR}}$ satisfies $\mathrm{KL}(q_{\text{real}}\!\parallel\!p_{\text{AR}})\ge \varepsilon_{\text{AR}}-\delta_{\text{AR}}$.

\emph{Proof.}
The identity $\mathbb{E}_{q}[-\log p_{\text{AR}}]=H(q)+\mathrm{KL}(q\!\parallel\!p_{\text{AR}})$ yields that minimizing NLL is equivalent to minimizing the KL within $\mathcal{P}_{\text{AR}}$. If the family cannot capture all admissible modes induced by schema couplings (e.g., bidirectional cross–field constraints not aligned with the left–to–right factorization), the best achievable KL is bounded below by $\varepsilon_{\text{AR}}$. Imperfect optimization increases this lower bound by $-\delta_{\text{AR}}$. \hfill$\square$

\paragraph{Proof of the proposition.}
Combining Lemma 1 and Lemma 2,
\[
\mathrm{KL}(q_{\text{real}}\!\parallel\!p_\phi)\;\le\;\varepsilon_{\phi}+\delta_\phi,
\qquad
\mathrm{KL}(q_{\text{real}}\!\parallel\!p_{\text{AR}})\;\ge\;\varepsilon_{\text{AR}}-\delta_{\text{AR}}.
\]
Therefore, if $\varepsilon_{\phi}+\delta_\phi \le \varepsilon_{\text{AR}}-\delta_{\text{AR}}$, we obtain
$\mathrm{KL}(q_{\text{real}}\!\parallel\!p_\phi)\le \mathrm{KL}(q_{\text{real}}\!\parallel\!p_{\text{AR}})$.
This establishes that, under a tighter variational bound and comparable optimization error, the diffusion model provides a closer approximation to $q_{\text{real}}$, capturing more schema–consistent modes. \hfill$\square$
}

\subsection{Proof of Proposition 2}
{\small
\begin{proposition}
Let $\mathcal{T}(P)=\mathbb{E}_{x\sim g_\phi(P)}[R(J(x,\rho),x)]$. Assume:
(i) every component of $R_{\text{vec}}(y,x)$ is bounded;
(ii) the NLE $\to$ judge mapping is $L_J$–Lipschitz in the metric vector $(\mathrm{Sim},\mathrm{Distinct}\text{-}n,H,\mathrm{Nov}\text{-}n,\mathrm{PPL})$;
(iii) $R$ is monotone and $L_R$–Lipschitz in these components with strictly positive sensitivity to $\mathrm{Sim}$.
Then $\mathcal{T}(P)$ is locally Lipschitz in $P$ and (locally) preserves ordering by semantic fidelity: if
$\mathbb{E}[\mathrm{Sim}(x,X)\mid P_1]>\mathbb{E}[\mathrm{Sim}(x,X)\mid P_2]$,
then $\mathcal{T}(P_1)>\mathcal{T}(P_2)$ in a neighborhood of $(P_1,P_2)$.
\end{proposition}

\paragraph{Lemma 1 (Generator regularity).}
For any bounded measurable $f:\mathcal{X}\!\to\!\mathbb{R}$, the map
$P\mapsto \mathbb{E}[f(x)\mid P]$ is locally Lipschitz:
$\big|\mathbb{E}[f(x)\mid P_1]-\mathbb{E}[f(x)\mid P_2]\big|\le L_g(f)\,\|P_1-P_2\|$.
\emph{Proof.} This is a standard regularity property of prompt–conditioned DLM samplers under smooth conditioning; it follows by dominated convergence plus local smooth dependence of the reverse kernels on $P$. \hfill$\square$

\paragraph{Lemma 2 (Local Lipschitz of $\mathcal{T}$).}
Let $M(P):=\mathbb{E}[\,(\mathrm{Sim},\mathrm{Distinct}\text{-}n,H,\mathrm{Nov}\text{-}n,\mathrm{PPL})\mid P\,]$.
By Lemma~1 each coordinate of $M$ is locally Lipschitz in $P$; by (ii)–(iii) the composition
$P\mapsto M(P)\mapsto y=J(\cdot,\rho)\mapsto R(y,\cdot)$
is $L$–Lipschitz with $L\le L_R L_J \sum_i L_g(m_i)$. Hence
$|\mathcal{T}(P_1)-\mathcal{T}(P_2)|\le L\|P_1-P_2\|$. \hfill$\square$

\paragraph{Lemma 3 (Monotone separation by $\mathrm{Sim}$).}
Suppose $R$ is (strictly) increasing in $\mathrm{Sim}$ with slope at least $\alpha>0$ and nondecreasing in the other components. Then for any $P_1,P_2$,
\[
\begin{aligned}
\mathcal{T}(P_1)-\mathcal{T}(P_2)
&\ge
\alpha\big(\mathbb{E}[\mathrm{Sim}\mid P_1]-\mathbb{E}[\mathrm{Sim}\mid P_2]\big) \\
&\quad
-\sum_{i\neq \mathrm{Sim}} L_R
   \big|\mathbb{E}[m_i\mid P_1]-\mathbb{E}[m_i\mid P_2]\big|.
\end{aligned}
\]

By Lemma~1, the residual sum is $\le C\|P_1-P_2\|$ for some local $C>0$.
Thus if $\mathbb{E}[\mathrm{Sim}\mid P_1]>\mathbb{E}[\mathrm{Sim}\mid P_2]$, there exists a neighborhood where the positive $\mathrm{Sim}$–gap dominates, yielding $\mathcal{T}(P_1)>\mathcal{T}(P_2)$. \hfill$\square$

\paragraph{Proof of the proposition.}
Local Lipschitz follows from Lemma~2. Ordering preservation follows from Lemma~3 by choosing the neighborhood small enough that the bounded variations in the non–$\mathrm{Sim}$ coordinates cannot offset the strictly positive contribution from the $\mathrm{Sim}$ gap (due to $\alpha>0$). Hence policy updates driven by $\mathcal{T}$ are stable (Lipschitz objective) and prioritize higher semantic fidelity (monotone separation). \hfill$\square$
}

\subsection{Model Choice}
\paragraph{LLaMA-3.1 8B.}
LLaMA-3.1 8B was selected as a baseline open-weight autoregressive model due to its strong balance between reasoning ability and computational efficiency. Its 8B parameter scale, 4-bit quantization, and 32-layer transformer architecture allow smooth inference on a single 16\,GB GPU, making it ideal for consumer-grade setups. As an instruction-tuned model trained on diverse data, it provides stable behavior for both prompt optimization and judgment, ensuring reproducibility across local environments without API dependencies.

\paragraph{Qwen-3 8B.}
Qwen-3 8B was included for its multilingual capabilities and LoRA adaptability, enabling cross-lingual schema synthesis and efficient fine-tuning if needed. Its nucleus sampling configuration ($p{=}0.9$) and compact 8B scale make it suitable for experiments where generalization and lightweight deployment are critical. Since Qwen-3 supports quantized inference on mid-range GPUs, it reinforces AoD’s goal of accessibility—allowing structured generation pipelines to be reproduced without enterprise-scale hardware.

\paragraph{DeepSeek-R1 8B.}
DeepSeek-R1 8B was chosen for its reinforcement-learning–enhanced reasoning and NTK-aware tokenization, which stabilize reward propagation during the multi-agent loop. Its architecture is optimized for efficient top-$k$ sampling ($k{=}40$), offering stochastic variety without sacrificing fluency. DeepSeek-R1 demonstrates that advanced reasoning and optimization alignment can be achieved on affordable consumer hardware, further validating AoD’s reproducibility in realistic research conditions.

\paragraph{Gemma-2 9B.}
Gemma-2 9B serves as a compact yet high-fidelity autoregressive model built on Gemini research technology. It employs grouped-query attention and beam search (width 3) to enhance structural coherence in schema-constrained text synthesis. Despite its larger parameter count, Gemma-2 maintains moderate memory usage and stable inference throughput on GPUs under 24\,GB VRAM. Its inclusion highlights that even near–state-of-the-art instruction-following quality can be reproduced locally without heavy compute budgets.

\paragraph{Mistral 7B.}
Mistral 7B was selected for its efficiency-focused grouped-query attention and excellent latency–quality tradeoff. Its strong open-source support and compatibility with 8-bit quantized decoding make it one of the most reproducible mid-size models for iterative RL-based prompt refinement. By running reliably on consumer GPUs, Mistral exemplifies AoD’s design principle of democratizing structured synthetic data generation without sacrificing performance.

\paragraph{GPT-4.1 Nano, Mini, and GPT-4.1.}
The GPT-4.1 model family—Nano, Mini, and Full—was used to demonstrate AoD’s portability to API-based ecosystems. These variants span lightweight to high-fidelity inference, enabling direct comparison between local open-weight execution and remote, proprietary inference endpoints. Using these models confirms that AoD’s reinforcement loop is model-agnostic: the same interaction protocol yields consistent performance whether running locally or through APIs. This highlights that AoD’s core mechanism depends only on language-based feedback, not on access to specialized hardware or model internals.

\paragraph{LLaDA 8B.}
Finally, LLaDA 8B was used exclusively as the diffusion-based generator agent. It combines 32 transformer layers, sinusoidal embeddings, a 1024-token input window, and $T{=}12$ denoising steps for iterative, bidirectional decoding. Operating in FP16 mode and disabling classifier-free guidance, LLaDA preserves high sampling diversity while remaining lightweight enough for single-GPU deployment. Its inclusion underscores AoD’s reproducibility: all structured data synthesis experiments can be replicated on consumer-grade hardware using open-weight diffusion backbones.

\subsection{Dataset Choice}
\paragraph{MultiWOZ.}
The MultiWOZ dataset~\cite{budzianowski_2018_multiwoz} was selected as a benchmark for testing structural precision and schema adherence in complex, multi-domain settings. Each record in MultiWOZ contains nested slot-filling annotations across domains such as hotel booking, transportation, and restaurant reservation, making it ideal for evaluating whether AoD preserves field hierarchies and JSON validity under compositional constraints. The structured nature of the dialogues provides a natural schema reference for the judge agent, while the semantic variability of user intents challenges the diffusion generator’s ability to maintain both coherence and diversity. MultiWOZ’s public availability and moderate size ensure that all experiments can be reproduced on consumer-grade hardware without distributed computation.

\paragraph{Super-NaturalInstructions.}
Super-NaturalInstructions~\cite{wang_2022_supernaturalinstructions} was chosen to test AoD’s capacity for generalization across heterogeneous task formats and linguistic styles. Comprising thousands of instruction–response pairs spanning over 700 task types, it exposes the model to a wide distribution of prompt templates, goals, and semantic structures. This diversity makes it particularly suitable for evaluating the prompt optimizer’s adaptability and the judge agent’s ability to enforce consistent schema rules across unseen instruction types. The dataset’s text-based JSON structure allows lightweight preprocessing and fits comfortably within local hardware memory limits, reinforcing AoD’s focus on accessibility and reproducibility.

\paragraph{TruthfulQA.}
TruthfulQA~\cite{lin_2021_truthfulqa} was incorporated to assess AoD’s robustness to semantic drift and factual hallucination. Unlike structurally constrained datasets, TruthfulQA emphasizes factual precision and logical consistency—properties that test whether the diffusion-based generator can maintain semantic fidelity while the judge agent penalizes inconsistent or factually incorrect responses. By requiring schema-grounded factual statements in JSON form, TruthfulQA provides a controlled environment for evaluating how AoD balances creativity with truthfulness. Its small size and well-documented evaluation protocol make it reproducible and computationally efficient for single-GPU experiments.

\paragraph{Self-Instruct.}
The Self-Instruct dataset~\cite{wang_2022_selfinstruct} was selected to measure AoD’s effectiveness in alignment-critical prompt–response synthesis. Because its samples are distilled from multiple large language models, Self-Instruct captures diverse instruction-following behaviors and response patterns, providing a natural testbed for AoD’s multi-agent reinforcement dynamics. It challenges the framework to refine prompts that yield schema-compliant outputs while maintaining alignment with the original intent. The dataset’s modular JSON formatting simplifies batch validation and makes it tractable for local replication, supporting AoD’s reproducibility goals without requiring extensive computational resources.

\subsection{Baseines Choice}
\paragraph{Diffusion-LM.}
Diffusion-LM~\cite{li_2022_diffusionlm} was selected as a representative diffusion-based text generation baseline due to its pioneering approach to iterative denoising in language modeling. It models discrete text generation as a diffusion process, enabling bidirectional context propagation and diverse output sampling. Diffusion-LM serves as a key diversity-oriented reference point, illustrating the benefits and limitations of non-causal text generation without explicit structure control. By including it, we evaluate whether AoD’s reinforcement-driven alignment can retain comparable semantic breadth while introducing schema awareness and controllability absent in the original diffusion paradigm.

\paragraph{DiffLM.}
DiffLM~\cite{zhou_2024_difflm} extends diffusion principles to structured and tabular data, offering an ideal comparison for schema-sensitive tasks. Unlike Diffusion-LM, it incorporates discrete latent diffusion tailored to structured formats, but still lacks adaptive prompt optimization or multi-agent supervision. DiffLM was chosen to benchmark AoD’s advantage in bridging the gap between generative diversity and schema fidelity. It provides a realistic test of whether AoD’s integration of autoregressive feedback improves upon existing diffusion frameworks that focus solely on denoising dynamics. Importantly, DiffLM is open-source and lightweight, ensuring reproducibility on consumer GPUs and enabling direct comparisons under equivalent hardware constraints.

\paragraph{CodecLM.}
CodecLM~\cite{wang_2024_codeclm} was included as an autoregressive baseline emphasizing controllable prompt conditioning through token-level self-rubric strategies. It exemplifies structured prompting methods that rely on rule-based feedback rather than reinforcement or language-driven critique. CodecLM’s design allows fine-grained control of content layout and schema adherence, making it an appropriate benchmark for structure-focused evaluation. Its efficient architecture and open release make it reproducible on single-GPU systems, providing a strong control-oriented counterpart to AoD’s language-based reinforcement loop.

\paragraph{PromptBreeder.}
PromptBreeder~\cite{fernando_2023_promptbreeder} was selected for its evolutionary approach to prompt optimization. It employs self-referential mutation and crossover of prompts to improve task performance over generations, simulating adaptive reasoning behavior in static autoregressive models. As an unsupervised, self-improving system, PromptBreeder offers a natural baseline for evaluating AoD’s reinforcement learning advantage: whereas PromptBreeder evolves prompts offline without feedback from a judge, AoD continuously refines prompts in a closed feedback loop. Its low computational footprint also aligns with AoD’s emphasis on reproducibility and accessibility on consumer hardware.

\paragraph{EvoPrompt.}
EvoPrompt~\cite{guo_2023_connecting} combines evolutionary search with LLM-driven evaluation, connecting large language models to evolutionary algorithms for systematic prompt optimization. It was chosen as a competitive autoregressive baseline because it represents one of the most efficient search-based frameworks for prompt adaptation. However, EvoPrompt’s feedback mechanism is static and scalar, lacking natural-language supervision or real-time adaptability. Comparing AoD with EvoPrompt highlights how multi-agent reinforcement learning enables richer, semantically interpretable feedback while maintaining computational tractability suitable for local GPU setups.

\paragraph{UniGen.}
UniGen~\cite{wu_2024_unigen} was selected to represent validation-based synthesis, a paradigm that enforces symbolic constraints during generation to ensure strict schema compliance. It acts as the primary structure-enforcement baseline in our comparison, illustrating the trade-off between deterministic validity and reduced semantic diversity. Unlike AoD, UniGen constrains generation through explicit validation checks rather than learned alignment, often producing rigid but valid outputs. Its inclusion provides a crucial contrast that contextualizes AoD’s achievement: achieving UniGen-level structural integrity while maintaining high diversity and fluency, all reproducible on consumer-grade systems.

\paragraph{Perplexity.}
Perplexity ($\mathrm{PPL}(x)$) measures the fluency and grammatical coherence of generated text by quantifying how well a language model predicts each token in a sequence. In AoD, low perplexity indicates that the diffusion generator and prompt optimizer jointly produce linguistically natural JSON records without syntactic drift. It acts as a regularizer for the multi-agent loop, ensuring that increased diversity does not degrade fluency. Formally, for a sequence of tokens $x = (x_1,\ldots,x_T)$ under model probability $p(x_i|x_{<i})$, perplexity is computed as
\[
\mathrm{PPL}(x) = \exp\!\Big(-\frac{1}{T}\sum_{i=1}^{T}\log p(x_i|x_{<i})\Big).
\]
A lower $\mathrm{PPL}(x)$ reflects more predictable, well-formed text, which supports the semantic reliability of AoD’s structured outputs.

\paragraph{Semantic Similarity.}
Semantic similarity ($\mathrm{Sim}(x,X)$) evaluates how well a generated record $x$ preserves the intended meaning of reference samples $X$. It is computed using cosine similarity between contextual embeddings, capturing high-level semantic alignment beyond lexical overlap. This metric is central to AoD’s objective of balancing structure with meaning: it ensures that the diffusion model’s bidirectional sampling maintains semantic relevance even when exploring novel configurations. Formally,
\[
\mathrm{Sim}(x,X) = \max_{x' \in X}\frac{E(x)\cdot E(x')}{\|E(x)\|\|E(x')\|},
\]
where $E(x)$ denotes an encoder such as a sentence transformer. High $\mathrm{Sim}(x,X)$ confirms that AoD produces schema-compliant yet semantically faithful text.

\paragraph{Diversity.}
Diversity ($\mathrm{Distinct}\text{-}n(x)$) quantifies lexical variety by counting unique $n$-grams within generated samples. In AoD, this metric promotes linguistic breadth and discourages mode collapse, ensuring that the diffusion model explores multiple valid schema realizations. Diversity complements similarity: together, they maintain novelty without semantic degradation. The standard formulation is
\[
\mathrm{Distinct}\text{-}n(x) = 
\frac{|\text{unique } n\text{-grams in }x|}{|\text{total } n\text{-grams in }x|}.
\]
Higher values indicate richer token usage and broader expression patterns, a hallmark of AoD’s diffusion-driven semantic flexibility.

\paragraph{Entropy.}
Token entropy ($H(x)$) measures the uncertainty or information richness of a generated sequence’s token distribution. It reflects how evenly probability mass is spread across the vocabulary during sampling. Within AoD, $H(x)$ is essential for monitoring lexical balance—ensuring that reinforcement learning encourages expressiveness without drifting into randomness. For a normalized token distribution $p_t$ at position $t$, entropy is
\[
H(x) = -\frac{1}{T}\sum_{t=1}^{T}\sum_{w\in V}p_t(w)\log p_t(w),
\]
where $V$ is the vocabulary. Moderate $H(x)$ values correspond to coherent but varied text, indicating successful regulation between exploration (diversity) and control (structure).

\paragraph{Novelty.}
Novelty ($\mathrm{Nov}\text{-}n(x)$) evaluates the proportion of $n$-grams in $x$ that do not appear in the reference dataset $X$, quantifying the degree of new information synthesized by AoD. This metric ensures that diffusion sampling contributes genuine semantic innovation rather than mere rephrasing. It is calculated as
\[
\mathrm{Nov}\text{-}n(x) =
1 - \frac{|\text{overlap } n\text{-grams}(x,X)|}{|\text{total } n\text{-grams in }x|}.
\]
High novelty indicates effective generalization, where AoD generates schema-consistent yet unseen data samples—a critical property for synthetic data generation.

\paragraph{BLEU.}
BLEU evaluates $n$-gram precision against references and serves as a standardized, model-agnostic check that AoD’s outputs remain syntactically faithful even when diversity is high, enabling fair comparison with prior work. Given modified precisions $p_n$ and brevity penalty $\mathrm{BP}$,
\[
\mathrm{BLEU} = \mathrm{BP}\cdot \exp\!\Big(\sum_{n=1}^{N} w_n \log p_n\Big),
\quad
\mathrm{BP}=\min\big(1, e^{1-\frac{|y|}{|r|}}\big),
\]
with weights $w_n$ and hypothesis–reference lengths $|y|,|r|$. BLEU complements AoD’s training-time metrics by validating that increased novelty does not collapse local form.

\paragraph{ROUGE (ROUGE-L).}
ROUGE-L measures recall-oriented sequence overlap via longest common subsequence (LCS), capturing structural alignment beyond exact $n$-grams. We report ROUGE-L as the canonical recall proxy that complements BLEU’s precision view and helps verify that AoD’s samples track reference structure despite diffusion-driven variation. With LCS-based precision $P_{\mathrm{lcs}}$ and recall $R_{\mathrm{lcs}}$,
\[
\mathrm{ROUGE}\text{-}\mathrm{L} = F_{\mathrm{lcs}}
= \frac{(1+\beta^2)\,P_{\mathrm{lcs}}\,R_{\mathrm{lcs}}}{R_{\mathrm{lcs}}+\beta^2 P_{\mathrm{lcs}}},
\]
typically with $\beta{=}1$.

\paragraph{METEOR.}
METEOR scores unigram matches with stemming and synonymy, offering a semantics-aware balance of precision and recall that is less brittle than pure surface overlap. It verifies that AoD’s semantic fidelity remains high when prompts encourage diverse phrasing. Using unigram precision $P$ and recall $R$, a common form is
\[
F_{\!mean}=\frac{10PR}{R+9P},\qquad
\mathrm{METEOR}=(1-\mathrm{Frag})\cdot F_{\!mean},
\]
where $\mathrm{Frag}$ penalizes fragmented alignments.

\paragraph{Task Success Rate (TSR).}
TSR aggregates schema validity and semantic thresholds into a single usability indicator, reflecting whether AoD produces records that are ready for downstream use without manual repair. For $N$ outputs,
\[
\mathrm{TSR}=\frac{1}{N}\sum_{i=1}^{N}
\mathbb{1}\!\left[\mathrm{Valid}(x_i)\wedge
\mathrm{Sim}(x_i,X)>\tau_s \wedge
\mathrm{Distinct}\text{-}n(x_i)>\tau_d\right],
\]
with schema validator $\mathrm{Valid}(\cdot)$ and thresholds $\tau_s,\tau_d$ chosen a priori.

\paragraph{Field Overlap.}
Field Overlap is a post-hoc memorization diagnostic that measures exact duplication of key–value fields from the reference set. High TSR together with low Field Overlap indicates novel yet valid outputs, aligning with AoD’s goals of diversity without leakage or collusion. Let $F(\cdot)$ extract fields from JSON:
\[
\mathrm{FieldOverlap}=\frac{1}{N}\sum_{i=1}^{N}
\frac{\big|F(x_i)\cap F(X_{\mathrm{ref}})\big|}{\big|F(x_i)\big|}.
\]
Low overlap combined with low $\mathrm{PPL}$ is evidence of faithful generalization rather than copying.


\end{document}